\documentclass[aps,prd,superscriptaddress,nofootinbib,amsmath,amsfonts,preprintnumbers,groupedaddress,showpacs,10pt,english]{revtex4-1}
\usepackage{amsmath}
\usepackage{amssymb}
\usepackage{babel}
\usepackage{wrapfig}
\usepackage{cancel}

\newcommand{\e}{\mathrm{e}}
\usepackage{relsize,exscale}
\makeatletter

\usepackage{array,multirow,graphicx}
\usepackage{dcolumn}
\usepackage{newlfont}
\usepackage{bm}
\usepackage[colorlinks,citecolor=blue,urlcolor=blue,linkcolor=blue]{hyperref}
\usepackage[figtopcap]{subfigure}
\usepackage{color}

\newcommand\be{\begin{equation}}
\newcommand\ba{\begin{eqnarray}}
\newcommand\ee{\end{equation}}
\newcommand\ea{\end{eqnarray}}

\allowdisplaybreaks[4]

\begin{document}
\tolerance=5000

\date{\today}
\title{Slow-rotating black holes with potential \\
in dynamical Chern-Simons modified gravitational theory}

\author{G.~G.~L.~Nashed$^1$}\email{nashed@bue.edu.eg}
\author{Shin'ichi~Nojiri$^{2,3}$}\email{nojiri@gravity.phys.nagoya-u.ac.jp}
\affiliation {$^1$ Centre for Theoretical Physics, The British University, P.O. Box
43, El Sherouk City, Cairo 11837, Egypt \\
$^2$ Department of Physics, Nagoya University, Nagoya 464-8602,
Japan \\
$^3$ Kobayashi-Maskawa Institute for the Origin of Particles and the Universe,
Nagoya University, Nagoya 464-8602, Japan }
\begin{abstract}

The Chern-Simons amended gravity theory appears as a low-energy effective theory of string theory.
The effective theory includes an anomaly-cancelation correction to the Einstein-Hilbert action.
The Chern-Simons expression consists of the product $\varphi R \tilde R $ of the Pontryagin density $R \tilde R $ with a scalar field $\varphi$,
where the latter is considered a background field
(dynamical construction or non-dynamical construction).
Many different solutions to Einstein's general relativity continue to be valid in the amended theories.
The Kerr metric is, however, considered an exceptional case that raised a search for rotating black hole solutions.
We generalize the solution presented in Phys. Rev. D \textbf{77}, 064007 (2008) by allowing the potential $V$ to have a non-vanishing value,
and we discuss three different cases of the potential, that is, $V=\mathrm{const.}$, $V\propto \varphi$, and $V\propto \varphi^2$ cases.
This study presents, for the first time, novel solutions prescribing rotating black holes in the frame of the dynamical formulation of the Chern-Simons gravity,
where we include a potential and generalize the previously derived solutions.
We derive solutions in the slow-rotation limit, where we write the parameter of the slow-rotation expansion by $\varepsilon$.
These solutions are axisymmetric and stationary, and they make a distortion of the Kerr solution by a dipole scalar field.
Furthermore, we investigate that the correction to the metric behaves in the inverse of the fourth order of radial distance from the center
of the black hole as $V\propto \varphi$.
This suggests that any meaningful limits from the weak-field experiments could be passed.
We show that the energy conditions associated with the scalar field of the case $V\propto \varphi$ are non-trivial and have non-trivial values to the leading order.
These non-trivial values come mainly from the contribution of the potential. Finally, we derived the stability condition using the geodesic deviations.
We conclude this study by showing that other choices of the potential, i.e., $V\propto \varphi^n$, where $n>2$ are not allowed
because all the solutions to these cases will be of order $\mathcal{O}\left(\varepsilon^2\right)$, which is not covered in this study.

\end{abstract}





\maketitle

\section{Introduction}\label{intro}

The goal of the modified gravitational theories is to deeply understand the physical phenomena that are difficult to explain in the frame of general relativity. Recently, different modified gravitational theories have been constructed to overcome one or more open issues in astrophysics and cosmology \cite{Sotiriou:2008rp, Nojiri:2010wj, Cai:2015emx,Nojiri:2017ncd, Nashed:2018cth,Nashed:2016tbj,Krssak:2018ywd,Olmo:2019flu, Cabral:2020fax,Harko:2020ibn, Capozziello:2021krv,Fernandes:2022zrq}.
When considering  black hole physics, a particular class of amended gravitational theories may yield a non-trivial mission.
We know that the Kerr and Schwarzschild black hole solutions are already standard for many different constructions,
and only dynamical behavior can supply robust tests for observation constraints. Therefore it could be much better if we prescribe discrepancies from general relativity by an appropriate set of parameters
and measure the deviation for the metric from the known black hole solutions and requiring that   Schwarzschild and Kerr solutions could be recovered in some limit.

This idea  was explained in \cite{Johannsen:2011dh}, where amendments to the Kerr black hole solution are executed naturally by using arbitrary values of spin,
through the use of axisymmetric, asymptotically flat, and stationary spacetime that describes a general rotating black hole solution in amended gravitational theory,
regardless theory creating them.
The output of this process was to overcome many of the issues that usually appear in perturbational theory in the frame of general relativity  because of
the violation of no-hair theorems \cite{Gair:2007kr, Johannsen:2010xs, Sotiriou:2015pka, Herdeiro:2014goa, Herdeiro:2015waa, Cardoso:2016ryw},
that strictly limit the predictions to certain cases, for example, the motion of stars and pulsars around black holes \cite{Wex:1998wt, Will:2007pp, Nashed:2021sji,Broderick:2013rlq}
or extreme mass-ratio inspiraling \cite{Barack:2006pq}.
The non-existence of a specific model accountable for the presence of the metric constructed in \cite{Johannsen:2011dh}, precludes its expansion to
other different settings than black hole solutions, raising the problem of how such types of solutions could be derived.

In this regard, the Chern-Simons gravity is a modified gravity and interesting theory capable of discussing quantum gravity and black hole issues
in an exclusive theoretical setting.
We should note that Jackiw and Pi first proposed the Chern-Simons gravity \cite{Jackiw:2003pm} motivated by the amendment given by the Chern-Simons
terms of electrodynamics \cite{Carroll:1989vb}.
In the frame of $U(1)$ gauge theory, the Lagrangian of the Maxwell theory is amended by including a term where a (pseudo)-scalar field $\varphi(x)$
couples to the $U(1)$-gauge topological Pontryagin density, $F^{\mu\nu}\tilde F_{\mu\nu}$, which is a pseudo scalar quantity and therefore parity odd.
Moreover, within the gauge invariance, such an amendment allows for the Lorentz and/or Parity symmetry violations \cite{Carroll:1989vb}.
The violations are generated by the amended term of the Carroll-Field-Jackiw expression i.e., $f_\mu \tilde F^{\mu\nu}A_{\nu}$
with $f_\mu \equiv \partial_\mu \varphi$ being the axial vector that is responsible for Lorentz and/or Parity symmetry breaking and $A_\nu$ is
the gauge potential of the electromagnetic field.
In general, $f_\mu $ links to one of the coefficients of the Lorentz and/or Parity violation in the standard model extension \cite{Colladay:1996iz,Colladay:1998fq,Cisterna:2018jsx,Corral:2021tww,Chatzifotis:2022mob,Chatzifotis:2022ene,Kostelecky:2003fs}.
In the same manner, the Chern-Simons amendment of the Maxwell electrodynamics adds to the action of general relativity with a non-minimal coupling between $\varphi(x)$
and the gravitational Pontryagin density which is figured as $\varphi R \tilde R$ where $R \tilde R\equiv R^{abcd} \tilde R_{abcd}$.

{ The Chern-Simons theory has been classified into two types: non-dynamical and dynamic.
In the non-dynamical case, the action lacks the scalar field kinetic term and therefore yields an extra limit on
the solution, i.e. the vanishing of the Pontryagin
density \cite{Grumiller:2007rv}. Also, in the non-dynamical frame, the impact of  Chern-Simons comes from the expression coded by the C-tensor \cite{Jackiw:2003pm, Alexander:2009tp,Konno:2009kg,Konno:2014qua}.
Moreover,  modified field equations generate the constraint, $R\tilde R=0$, to ensure the diffeomorphism invariance of the theory. However, in the dynamical case,  the kinetic term of the scalar field is kept in action, and the scalar field is governed by a wave equation whose source is the Pontryagin density. Thus,
the dynamical case keeps diffeomorphism invariance and the strong equivalence principle, in spite that it violates GR's Birkhoff theorem \cite{Grumiller:2007rv,Yunes:2007ss} and the principle of effacement
principle \cite{Yagi:2011xp}.}

The importance of the Chern-Simons gravity theory when we discuss the problem presented in \cite{Johannsen:2010xs}, can then be estimated by considering
the properties of the symmetry for the Pontryagin density under parity transformations.
The necessities of the Chern-Simons expression to keep parity yields the scalar coupling where a pseudo-scalar field (parity odd) mediates,
which could be consistent with the hypothetical string theory.
This tells that parity-even solutions, like the Schwarzschild solution which has spherical symmetry, are not impacted by the Chern-Simons corrections but
the impacts appear only in scenarios that violate parity, i.e., in the Kerr solution or general rotating black holes solutions \cite{Yunes:2009hc}.
In another meaning, the Chern-Simons term can automatically generate the distortions of the Kerr solution and provide a theoretical justification \cite{Johannsen:2010xs}. {  Therefore, it becames clear that the remarkable effect of the Chern-Simons expression in these solutions was fundamental in finding a large set of totally causal solutions
in a manner characteristic of general relativity.}
Moreover, many novel black holes have been derived, like rotating  G\"{o}del-type spacetimes \cite{Porfirio:2016nzr,Porfirio:2016ssx,Agudelo:2016pic,Altschul:2021rog}
or the Einstein-dilaton Gauss-Bonnet gravity
\cite{Kanti:1995vq,Kanti:1997br,Kleihaus:2011tg,Ayzenberg:2014aka,Maselli:2015tta,Kleihaus:2015aje,Okounkova:2019zep,Cano:2019ore,Delgado:2020rev,Pierini:2021jxd}.

The role of the Chern-Simons term in the framework of amended gravitational theories is also robustly dealt with
by different justifications originating from various physical backgrounds, in which the presence
of the Chern-Simons expression appears to be diffuse \cite{Alexander:2009tp}.
In the frame of particle physics, the gravitational anomaly becomes proportionate to the Pontryagin density,
and the corresponding Chern-Simons-like term should be involved in the Lagrangian to remove the oddity.
The corresponding term of such type can be created in string theory through the Green-Schwarz technique and appear
in low-energy string models \cite{Smith:2007jm, Adak:2008yg}.
Wonderfully, some similarities can be summarized with loop quantum gravity methods \cite{Ashtekar:1988sw} where the Chern-Simons expression arises in discussing the chiral anomaly of fermions and the Immirzi field ambiguity \cite{Perez:2005pm,Freidel:2005sn,Date:2008rb,Mercuri:2009zi,Mercuri:2009vk}.
Additionally, such theories may guide in calculating new techniques to investigate the local Lorentz and/or Parity symmetry breaking in gravitation,
that is forecasted to gain new observational results soon.
The Chern-Simons parity violation impacts are already well discussed in contexts like birefringence with amplitude gravitational
wave \cite{Jackiw:2003pm,Martin-Ruiz:2017cjt,Nojiri:2019nar,Nojiri:2020pqr},
the baryon asymmetry problem \cite{Alexander:2004us,Garcia-Bellido:2003wva,Alexander:2004xd}
and CMB polarization \cite{Alexander:2006mt,Lue:1998mq,Bartolo:2018elp,Bartolo:2017szm}.
Moreover, the effect of the Chern-Simons term has been discussed in the frame of metric affine theory \cite{Hehl:1994ue, Zanelli:2005sa, Boudet:2022wmb}
and in the frame of Cartan formalism \cite{Hehl:1990ir,Banados:2001xw,Cacciatori:2005wz,BottaCantcheff:2008pii}.

In the frame of the dynamical formalism of the Chern-Simons gravity, Yunes and Pretorius \cite{Yunes:2007ss} have derived a non-trivial solution
to keep the parity violation and neglect the potential\footnote{In this study,
we follow the terminology presented in \cite{Yunes:2007ss}.}.
The present research aims to derive a new weakly rotating black hole solution taking into account the potential.
For this purpose, we will discuss three different cases of the potentials that affect the scalar field and the equations of motion.
The outline of the present paper is as follows,
In Sec.~\ref{ABC}, we present the ingredients of Chern-Simons gravity theory.
In Sec.~\ref{axisym}, we apply the field equations of the dynamical models that include the potential to a line element describing
a slowly rotating black hole which is true for the small Chern-Simons coupling constants.
We try to understand some of the related physics of the derived solutions by calculating their geodesics deviation and stating their stability condition in Sec.~\ref{properties}.
In Sec.~\ref{conclusions}, we discuss the main results of the present study and give possible future work.

The following conventions are used throughout the present study:
We use four-dimensional spacetimes that have the following signature $(-,+,+,+)$~\cite{Misner:1973prb}, square and round square parentheses refer to
anti-symmetrization and symmetrization respectively, i.e., $T_{[ab]}=\frac{1}{2} \left(T_{ab}-T_{ba} \right)$ and $T_{(ab)}=\frac{1}{2} \left( T_{ab}+T_{ba} \right)$.
The partial derivatives are refereed by commas $\left(\mbox{e.g.,}\ \frac{\partial\varphi}{\partial r}=\partial_r\varphi=\varphi_{,r} \right)$.
The Einstein summation is applied 
and we use the unit where the light speed $c$ is unity, $c=1$.

\section{Chern-Simons modified gravity}
\label{ABC}

In the present section, we discuss the relevant topics which yield a total form of the Chern-Simons amended gravity and state some notation
\cite{Alexander:2009tp}.

\subsection{Basics of Chern-Simons gravity}

The action of the Chern-Simons gravity theory is defined as,
\begin{align}
\label{Chern-Simonsaction}
S = S_\mathrm{EH} + S_\mathrm{CS} + S_{\varphi} + S_\mathrm{matter}\, ,
\end{align}
with
\begin{align}
\label{EH-action}
S_\mathrm{EH} =&\, \kappa \int_V d^4x \sqrt{-g} R\, , \\
\label{Chern-Simons-action}
S_\mathrm{CS} =&\, \frac\mu {4} \int_V d^4x \sqrt{-g} \varphi R \tilde R \,, \\
\label{Theta-action}
S_{\varphi} =&\, - \frac{\nu }{2} \int_V d^4x \sqrt{-g} \left[ g^{ab}
\left(\nabla_{a} \varphi\right) \left(\nabla_{b} \varphi\right) + 2 V(\varphi) \right]\, , \\
S_\mathrm{matter} =&\, \int_V d^4x \sqrt{-g} L_\mathrm{matter}\, .
\end{align}
In Eq.~\eqref{Chern-Simonsaction}, $S_\mathrm{EH}$ is the Einstein-Hilbert action, $S_\mathrm{CS}$ is the Chern-Simons rectification,
$S_{\varphi}$ represents the kinetic term, and the potential of the (pseudo) scalar-field $\varphi$ and $S_\mathrm{matter}$ expresses the action
of matters, where $L_\mathrm{matter}$ is the matter Lagrangian density.
The following conventions will be used throughout this study: $\kappa = \frac{1}{16 \pi G}$,
$g$ is the determinant of the metric, $\mu$ and $\nu$ are dimensional constants, $\nabla_a$ is the covariant derivative,
$R$ is the Ricci scalar curvature, and $a,b,c,\cdots=0,1,2,3$.
The expression $R \tilde R $ is the Pontryagin density, figured as,
\begin{align}
\label{pontryagindef}
R \tilde R = {R^b}_{acd} \tilde R^{a\ cd}_{\ b} \,,
\end{align}
with $\tilde R^{a\ cd}_{\ b}$ being the dual Riemann-tensor defined by
\begin{align}
\label{Rdual}
\tilde R^{a\ cd}_{\ b} =\frac{1}{2} \epsilon^{cdef}{R^a}_{bef}\,,
\end{align}
where $\epsilon^{abcd}$ is the 4-dimensional Levi-Civita tensor which is a completely skew-symmetric tensor with $\epsilon^{0123}=-1$.

The Chern-Simons scalar field $\varphi$ is a function of the spacetime coordinates and $\varphi$ parametrizes deviation from general relativity.
When $\varphi = \textrm{const.}$, the Chern-Simons gravity theory coincides with Einstein's general relativity theory since the Pontryagin density is the total divergence
of the Chern-Simons topological current $\mathcal{K}^a$
\begin{align}
\nabla_a \mathcal{K}^a = \frac{1}{2} R \tilde R \, .
\label{eq:curr1}
\end{align}
Here
\begin{align}
\mathcal{K}^a =\epsilon^{abcd} \Gamma^e_{bf} \left(\partial_{c}\Gamma^f_{de}+\frac{2}{3} \Gamma^f_{cg}\Gamma^g_{de}\right)\,,
\label{eq:curr2}
\end{align}
and $\Gamma^a_{bc}$ is the Christoffel second kind connection.
The use of Eq.~\eqref{eq:curr2} enables us to rewrite $S_\mathrm{CS}$ in the form \cite{Yunes:2007ss},
\begin{align}
\label{Chern-Simons-action-K}
S_\mathrm{CS} = \mu \left. \left( \varphi \; \mathcal{K}^a \right) \right|_{\partial {{V}}} - \frac\mu {2} \int_V d^4x \sqrt{-g} \;
\left(\nabla_{a} \varphi \right) \mathcal{K}^a\,.
\end{align}
The first expression of Eq.~\eqref{Chern-Simons-action-K} is usually omitted because it is calculated on the boundary of the spacetime~\cite{Grumiller:2008ie}, whilst the second expression is the Chern-Simons correction.

The equation of motions of the action \eqref{Chern-Simonsaction} can be derived by variation w.r.t. the metric and to the Chern-Simons coupling scalar field $\varphi$ that yield,
\begin{align}
\label{eom}
R_{ab} + \frac\mu {\kappa} C_{ab} =&\, \frac{1}{2 \kappa} \left(T_{ab} - \frac{1}{2} g_{ab} T \right)\, , \\
\label{eq:constraint}
\nu \square \varphi =&\, \nu \; \frac{dV}{d\varphi} - \frac\mu {4} R \tilde R \, .
\end{align}
Here $R_{ab}$ is the Ricci tensor, $\square \equiv \nabla_\mu \nabla^\mu$ is the D'Alembertian, and
$C_{ab}$ is the C-tensor defined as,
\begin{align}
\label{Ctensor}
C^{ab} = f_c \epsilon^{cde(a}\nabla_eR^{b)}{}_d+f_{cd}\tilde R^{d(ab)c}\,,
\end{align}
where
\begin{align}
\label{v}
f_a=\nabla_a\varphi\,,\quad
f_{ab}=\nabla_a\nabla_b\varphi\, .
\end{align}
Finally, the total stress-energy tensor $T_{ab}$ is defined as,
\begin{align}\label{Tab-total}
T_{ab} = T^\mathrm{matter}_{ab} + T_{ab}^{\varphi}\, ,
\end{align}
where $T^\mathrm{matter}_{ab}$ is the matter energy-momentum tensor (which we will set equal to zero in this study),
and $T_{ab}^{\varphi}$ is the energy-momentum tensor for the scalar field $\varphi$, which is defined as,
\begin{align}
\label{Tab-theta}
T_{ab}^{\varphi}
= {\nu} \left[ \left(\nabla_{a} \varphi\right) \left(\nabla_{b} \varphi\right)
 - \frac{1}{2} g_{ab}\left(\nabla_{a} \varphi\right) \left(\nabla^a \varphi\right)
 - g_{ab} V(\varphi) \right]\, .
\end{align}

In the frame of the Chern-Simons gravity, the strong equivalence principle, $\nabla_{a} T^{\mathrm{matter}\, ab} = 0$,
is satisfied assuming that Eq.~\eqref{eq:constraint} for the scalar field $\varphi$ hold.
We can show $\nabla_{a} T^{\mathrm{matter}\, ab} = 0$ by using the Bianchi identities for Eq.~\eqref{eom} and the equation,
\begin{align}
\label{nablaC}
\nabla_a C^{ab} = - \frac{1}{8} f^b R \tilde R \, .
\end{align}
The equality of Eq.~\eqref{nablaC} yields Eq.~\eqref{eq:constraint}.

{
Under the parity transformation, the Pontryagin density $R \tilde R $ changes its signature $R \tilde R \to - R \tilde R $.
Then if the scalar field $\varphi$ is the pseudo scalar, which changes its signature under the parity transformation $\varphi\to - \varphi$,
the Chern-Simons term $S_\mathrm{CS}$ (\ref{Chern-Simons-action}) is invariant under the parity transformation.
Therefore if the potential $V(\varphi)$ in (\ref{Theta-action}) is an even function of $\varphi$, $V(-\varphi)=V(\varphi)$, the model has the parity symmetry.
This tells that if there is a solution, we find another solution by changing the signature of the Pontryagin density $R \tilde R $ and the pseudo scalar $\varphi$,
$R \tilde R \to - R \tilde R $ and $\varphi\to -\varphi$, which corresponds to the time reversal $t\to t$ or spatial inversion $\bm{r}\to - \bm{r}$.
On the other hand, if the potential $V(\varphi)$ in (\ref{Theta-action}) is not an even function of $\varphi$, that is, $V(-\varphi)\neq V(\varphi)$,
the parity symmetry of the model is explicitly broken.
}

\subsection{Two constructions of Chern-Simons gravity}
\label{dyn-vs-can}

We can classify the Chern-Simons gravity into two different constructions:
The non-dynamical one and the dynamical one.
The non-dynamical construction is defined by setting $\nu = 0$~\footnote{Generally, $\mu = \kappa$ when we work in the non-dynamical construction,
This is not, however, important and we choose to set such constant to be arbitrary.}, in which the field equations yield,
\begin{align}
\label{eom-nd}
R_{ab} + \frac\mu {\kappa} C_{ab} =&\, \frac{1}{2 \kappa} \left(T_{ab}^\mathrm{matter}- \frac{1}{2} g_{ab} T^\mathrm{matter} \right)\, , \\
\label{eq:constraint-nd}
0 =&\, R \tilde R \, .
\end{align}
If we consider the the vacuum, we find $T_{ab}^\mathrm{matter} = T^\mathrm{matter} =0$ in Eq.~\eqref{eom-nd}.
On the other hand, Eq.~\eqref{eq:constraint-nd} is called the Pontryagin constraint, which is, however,
used to be an evolution equation for the scalar field $\varphi$.

In the frame of the non-dynamical construction, the Pontryagin constraint not only reduces the space of allowed solutions but one should describe in advance
the entire history of the Chern-Simons coupling $\varphi$. If this description is supposed, then the Chern-Simons scalar field $\varphi$ is not affected by any interaction
and we may choose the so-called canonical choice of $\varphi$ given by
\begin{align}
\label{canonicalchoice}
\varphi_\mathrm{can} = \frac{t}{\alpha}\, ,
\quad
f^a_\mathrm{can} = \left[\frac{1}{\alpha},0,0,0\right]\, ,
\end{align}
which was supposed in the frame of non-dynamical case ~\cite{Jackiw:2003pm}.
The canonical choice (\ref{canonicalchoice}) simplifies the field equations but the choice is not truly ``canonical'' for the Chern-Simons scalar field $\varphi$.
In addition, the choice is not invariant under the coordinate transformation,
and therefore, the theory has no justification for why the special slicing of the manifold is chosen.
Such a choice is restrictive and does not allow for any axisymmetric rotating BH solutions
in the Chern-Simons gravity~\cite{Alexander:2007zg, Alexander:2007vt, Konno:2007ze, Yunes:2007ss}.

For the dynamical construction, the parameter $\nu$ to be arbitrary, in which the amended equation of motions is given by
Eqs.~\eqref{eom}-\eqref{Tab-theta}.
Equation~\eqref{eq:constraint} becomes the evolution equation of the Chern-Simons scalar field $\varphi$.
Thus, no constraint is put on the allowed space.
Instead of describing the history of the Chern-Simons field $\varphi$,
one needs to fix initial conditions for $\varphi$, which evolves in a self-consistent way through Eq.~\eqref{eq:constraint}.

The non-dynamical and dynamical constructions constitute unequal theories, despite sharing some properties in the action.
Despite that, it is allowed to take the limit $\nu \to 0$ in the action to derive the non-dynamical
frame, however, it is not expected that the same process works properly on the solutions of the dynamic case
so that the solutions to the non-dynamical frame could be recovered.

Before we end this section, we discuss the dimensions of constants and scalar field.
The fixation of one of the units $(\mu,\nu,\varphi)$ will limit the units of the others.
As an example, if the Chern-Simons scalar field have the dimension $[\varphi] = l^a$, then $[\mu] = l^{2 - a}$ and $[\nu] = l^{-2a}$, with $l$ being a length unit.
In a natural choice, the Chern-Simons scalar $\varphi$ could be dimensionless as in the standard scalar-tensor theories, where we find $[\mu] = l^2$
and $\nu$ be dimensionless quantity\footnote{Here we are going to put $G = c = 1$, and therefore, the units of the action becomes $l^2$.
If we put $h = c = 1$, then the unit of the action will be dimensionless and if $[\varphi] = l^a$ thence $[\mu] = l^{-a}$ and $[\nu] = l^{-2 a - 2}$.}.
Other choice is to put $\mu = \nu$, therefore putting $S_{\varphi}$ and the action of $S_\mathrm{CS}$ on equal footing; we then have $[\varphi] = l^{-2}$.
No construction demands us to choose particular units for $\varphi$, thus we will allow these to be arbitrary, as the results of the previous studies choose a different selection.

\section{Rotating black hole solutions with the non-trivial value of the potential in dynamical Chern-Simons gravity }\label{axisym}

In this section, we consider the rotating BHs in the dynamical formulation by taking into account the potential.
It is a very fictitious mission to study the stationary axisymmetric spacetime in the framework of the Chern-Simons gravity without making any approximation for the calculation.
Thus we are going to use a couple of approximations.
Then, we proceed to solve the amended Chern-Simons field equations to the second order in the perturbative expansion.

\subsection{The approximation process}
\label{approx}

Now we are going to use two approximation schemes: Small-coupling and slow-rotation approximations.
The small-coupling process treats the Chern-Simons amended gravity as a small deformation of Einstein's general relativity, which permits
{ to use of metric decomposition (including the second order) given by \cite{Yunes:2009hc}},
\begin{align}
g_{a{a_1}} = g_{a{a_1}}^{(0)} + \xi g^{(1)}_{a{a_1}}(\varphi) + \xi^2 g^{(2)}_{a{a_1}}(\varphi)\,,
\label{small-cou-exp0}
\end{align}
where $g_{a{a_1}}^{(0)}$ is the background metric satisfying the Einstein equations, such as the Kerr metric, whilst $g_{a{a_1}}^{(1)}(\varphi)$ and $g_{a{a_1}}^{(2)}(\varphi)$
are the first and second-order perturbations for the Chern-Simons gravity that depend on the scalar field $\varphi$.
The parameter $\xi$ stands for the small-coupling given by $\varphi=\mathcal{O}(\xi)$.

The slow-rotation process requires to re-expand the background metric in addition to the $\xi$-perturbations in the powers
of $a_\mathrm{Kerr}$.
{ Then the background and the metric perturbation are given by \cite{Yunes:2009hc}},
\begin{align}
\label{small-cou-exp}
g_{a{a_1}}^{(0)} =&\, \eta_{a{a_1}}^{(0,0)} + \varepsilon h_{a{a_1}}^{(1,0)} + \varepsilon^2 h_{a{a_1}}^{(2,0)}\, ,
\nonumber \\
\xi g_{a{a_1}}^{(1)} =&\, {\color {red} \xi} h_{a{a_1}}^{(0,1)} + \xi \varepsilon h_{a{a_1}}^{(1,1)} + \xi \varepsilon^2 h_{a{a_1}}^{(2,1)}\, ,
\nonumber \\
\xi^2 g_{a{a_1}}^{(2)} =&\, \xi^2 h_{a{a_1}}^{(0,2)} + \xi^2 \varepsilon h_{a{a_1}}^{(1,2)} + \xi^2 \varepsilon^2 h_{a{a_1}}^{(2,2)}\, .
\end{align}
Here the parameter $\varepsilon$ is the parameter of the slow-rotation expansion, $\varepsilon = \mathcal{O}\left(a_\mathrm{Kerr}\right)$.
We must remember that the { indices} of the notation $h^{(a,{a_1})}_{m n}$ label for terms of $\mathcal{O}\left( a,{a_1} \right)$, which expresses a term of
$\mathcal{O} \left( \varepsilon^a \right)$ and $\mathcal{O} \left( \xi^{a_1} \right)$.
As an example, in Eq.~\eqref{small-cou-exp}, $\eta_{a{a_1}}^{(0,0)}$ is the background metric when the rotation parameter vanishing, i.e., $a_\mathrm{Kerr} = 0$,
whilst $h_{a{a_1}}^{(1,0)}$ and $h_{a{a_1}}^{(2,0)}$ are the first and the second-order perturbations of the background in the rotation $a_\mathrm{Kerr}$.

{ Combining the two expansion processes, we obtain the expressions in terms of $\xi$ and $\varepsilon$,
which yield \cite{Yunes:2009hc}},
\begin{align}
g_{a{a_1}} = \eta_{a{a_1}}^{(0,0)} + \varepsilon h_{a{a_1}}^{(1,0)} + \xi h_{a{a_1}}^{(0,1)} + \varepsilon \xi h_{a{a_1}}^{(1,1)} + \varepsilon^2 h_{a{a_1}}^{(2,0)} + \xi^2 h_{a{a_1}}^{(0,2)}\,.
\end{align}
The first-order terms, means the expressions of $\mathcal{O}(1,0)$ or $\mathcal{O}(0,1)$, whilst second-order expressions refer
to $\mathcal{O}(2,0)$, $\mathcal{O}(0,2)$ or $\mathcal{O}(1,1)$.

In this study, the slow-rotation process corresponds to the expansion of the Kerr parameter $a_\mathrm{Kerr}$, and therefore
the dimensionless expansion parameter should be $\frac{a_\mathrm{Kerr}}{M}$.
Thus, the equations multiplied by $\varepsilon^n$ is of order $\mathcal{O}\left( \left(\frac{a_\mathrm{Kerr}}{M} \right)^n\right)$.

\subsection{The slow rotating BH solutions}\label{slow-rot}

For the background metric, the slow-rotating expansion is formulated through the Hartle-Thorne~\cite{Thorne:1984mz, Hartle:1968si},
with the following line element:
\begin{align}
\label{slow-rot-ds2}
ds^2 = -A \left[1 + a(r,\theta)\right] dt^2
+ \frac{1}{A} \left[1 + b(r,\theta)\right] dr^2
+ r^2 \left[1 + d(r,\theta) \right] d\theta^2
+ r^2 \sin^2{\theta} \left[1 + f(r,\theta) \right] \left[ d\phi - \Omega(r,\theta) dt \right]^2,
\end{align}
with $A = 1 - \frac{2M}{r}$, which appears in the Schwarzschild solution, and $M$ is the mass of the black hole in the absence of the Chern-Simons term.
In Eq.~\eqref{slow-rot-ds2}, we use the Boyer-Lindquist coordinates, i.e., $(t,r,\theta,\phi)$ and the metric
{ perturbations are $a(r,\theta)$, $b(r,\theta)$, $d(r,\theta)$, $f(r,\theta)$, and $\Omega(r,\theta)$ \cite{Yunes:2009hc}}.
When the metric perturbations vanish, the Schwarzschild metric is recovered.

{
The combination of the spacial inversion $\bm{r}\to -\bm{r}$ and spacial rotation gives $z\to -z$, which corresponds to $\phi \to -\phi$
or $d\phi \to - d\phi$.
Therefore $\Omega(r,\theta)$ in the line element (\ref{slow-rot-ds2}) is relevant for the parity symmetry.
For the spacetime which is a solution, if the spacetime obtained by replacing $\varphi\to -\varphi$ and $\phi\to -\phi$ $\left( d\phi\to - d\phi\right)$
is also a solution, the parity symmetry of the model is not broken.
}

The metric~\eqref{slow-rot-ds2} is rewritten as in~\cite{Thorne:1984mz, Hartle:1968si}, however, the perturbations of the metric should
be expanded in a series in both $\xi$ and $\varepsilon$.
By keeping the second order, { we have \cite{Yunes:2009hc}},
\begin{align}
\label{cons}
a(r,\theta) =&\, \varepsilon a_{(1,0)} + \varepsilon \xi a_{(1,1)} + \varepsilon^2 a_{(2,0)}\, , \nonumber \\
b(r,\theta) =&\, \varepsilon b_{(1,0)} + \varepsilon \xi b_{(1,1)} + \varepsilon^2 b_{(2,0)}\, , \nonumber \\
d(r,\theta) =&\, \varepsilon d_{(1,0)} + \varepsilon \xi d_{(1,1)} + \varepsilon^2 d_{(2,0)}\, , \nonumber \\
f(r,\theta) =&\, \varepsilon f_{(1,0)} + \varepsilon \xi f_{(1,1)} + \varepsilon^2 f_{(2,0)}\, , \nonumber \\
\Omega(r,\theta) =&\, \varepsilon \Omega_{(1,0)} + \varepsilon \xi \Omega_{(1,1)} + \varepsilon^2 \Omega_{(2,0)}\, .
\end{align}
Equations~\eqref{cons} have no expressions of $\mathcal{O}(0,0)$ because those terms are already included in the Schwarzschild metric
when the metric perturbations vanish in Eq.~\eqref{slow-rot-ds2}.
Moreover, we suppose that we recover the Schwarzschild spacetime as a solution in the limit $a_\mathrm{Kerr} \rightarrow 0$.
This tells that all expressions of $\mathcal{O}(0, a)$ should vanish.
Therefore the Chern-Simons expression should be linear to the Kerr rotation parameter $a_\mathrm{Kerr}$. { What exactly are the  parameters $\epsilon$ and { $\xi$}? The slow-rotation procedure is an expansion of the Kerr parameter,  $a_{Kerr}$, and thus its dimensionless expansion parameter should be $J=a/M$. Thus, any expression in the equations multiplied by $\varepsilon^s$ is of ${\cal{O}}\left((a/M)^s\right)$. The small-coupling expansion parameter should depend on the ratio of Chern-Simons coupling to the GR coupling, i.e. $\alpha/\kappa$, since this combination multiplies the C-tensor given by  Eq.~\eqref{Ctensor}.  Equation \eqref{Ctensor}  states in a clear way that the C-tensor is proportional to the gradients of the scalar field of Chern-Simons, which should proportional to $\mu/\nu$ because of the $\varphi$-evolution equation, Eq. \eqref{eq:constraint}. This means  that the Chern-Simons correction to the metric will be proportional to  ${ \xi} = (\mu/\kappa) (\mu/\nu)$. This term  is not dimensionless, and therefore, it is not formally a perturbation parameter. Since the only mass scale is available at the background metric, which up to the first order in the slow-rotation expansion is the BH mass. Thus, we choose to normalize ${ \xi}$ so that the parameter is $\xi$ multiplies terms of ${\cal{O}}\left[\alpha^{2}/(\kappa \beta M^4)\right]$. }

By using the slow-rotation limit in the Kerr metric for general relativity, the metric perturbations proportional
to $\xi^{0}$ to the first order are given as \cite{Yunes:2007ss},
\begin{align}
a_{(1,0)} =&\, b_{(1,0)} = d_{(1,0)} = f_{(1,0)} =0\,, \quad
\Omega_{(1,0)}=\frac{2{ \varepsilon} M a_\mathrm{Kerr}}{r^3},
\end{align}
and to second order as,
\begin{align}
a_{(2,0)} =&\, \frac{2{ \varepsilon^2} {a_\mathrm{Kerr}}^2 M}{A r^3} \left( \cos^2{\theta} + \frac{2 M} r \sin^2{\theta} \right)\,,\quad
b_{(2,0)} =\frac{{{ \varepsilon}^2 a_\mathrm{Kerr}}^2}{r^2} \left( \cos^2{\theta} - \frac{1}{A} \right)\,, \quad
d_{(2,0)} = \frac{{{ \varepsilon^2} a_\mathrm{Kerr}}^2}{r^2} \cos^2{\theta}\,, \nonumber \\
f_{(2,0)} =&\, \frac{{{ \varepsilon^2} a_\mathrm{Kerr}}^2}{r^2} \left(1 + \frac{2 M} r \sin^2{\theta} \right)\, , \quad
\Omega_{(2,0)} = 0,\,.
\end{align}
All the fields are expanded by the parameters $\xi$ and $\varepsilon$ involving the Chern-Simons field.
To obtain the leading-order terms for $\varphi$, we should obtain the evolution equation, Eq.~\eqref{eq:constraint}.
 Eq.~\eqref{eq:constraint}, shows that $\partial^2 \varphi \sim \left( \mu/\nu \right) R \tilde R $, where
the Pontryagin density becomes null to the zeroth order in $\frac{a_\mathrm{Kerr}}{M}$.
Therefore, the first order of the Chern-Simons scalar field should be $\varphi \sim \left(\frac{\mu}{\nu}\right) \left(a_\mathrm{Kerr}{M}\right)$,
that is proportional to $\varepsilon$.
Moreover, under the assumption that the unique solution in the zeroth-angular momentum should be the Schwarzschild metric, we find $\varphi^{(0,n_1)} = 0$ for all $n_1$.
The study presented in \cite{Yunes:2007ss} was very interested to derive a deformation of the Kerr solution in the dynamical formulation
when the potential is absent.
In the present study, we expand the study in \cite{Yunes:2007ss} to include the potential and discuss three different cases of the potential $V$:
$V=\mathrm{const.}$, $V\propto \varphi$ and $V\propto \varphi^2$. The first case, $V=\mathrm{const.}$ will not affect Eq.~\eqref{eq:constraint} but effect Eq.~\eqref{eom}.
In the second case, $V\propto \varphi$, both Eqs.~\eqref{eq:constraint} and Eq.~\eqref{eom} are affected.
Finally, the third case, $V\propto \varphi^2$, Eq.~\eqref{eq:constraint} is effected and Eq.~\eqref{eom} is not affected.
Other choices of the potential is not allowed, i.e., $V \propto \varphi^n$, $n>2$ because it yields solutions of order $\mathcal{O} \left(\varepsilon^2\right)$.

Because we have not fixed the dimensions of the Chern-Simons scalar field $\varphi$, $[\varphi] = l^a$, in the first case $V=\mathrm{const.}$,
we put $V=1$ by choosing $a=1$ so that the dimension of the kinetic term of $\varphi$ is unity,
$\left[ g^{ab}\left(\nabla_{a} \varphi\right) \left(\nabla_{b} \varphi\right) \right] = 1$ in (\ref{Theta-action}).
Similarly in the second case $V\propto \varphi$, we put $V= \varphi$ by choosing $a=\frac{3}{2}$.
In the third case $V\propto \varphi^2$, however, we need to introduce a new parameter $m^2$, $V= m^2 \varphi^2$ with $[m]=l^{-1}$, and
$m$ can be identified with the mass of $\varphi$.

We should also note that in the first and third cases, the potential is an even function of $\varphi$ and therefore the parity symmetry is not broken
but in the second case, the symmetry is explicitly broken.

In the following subsections, we are going to discuss the three physical cases of the potential in detail.


\subsection{The case of $V=\mathrm{const.}$}

The results presented in \cite{Yunes:2007ss} will be identical with the results of the case $V=\mathrm{const.}$
{ Therefore, we will not go into details of this case.}
\subsection{The case of $V\propto \varphi$}\label{Vvarphi}

In this case, the parity symmetry is explicitly broken because $V(\varphi)$ is the odd function of $\varphi$.
The scalar field equation~\eqref{eq:constraint} will be affected because $dV=\mathrm{const.}$.
By applying Eq.~\eqref{eq:constraint} to the line element \eqref{slow-rot-ds2} and by using Eq.~\eqref{cons}, we obtain,
\begin{align}
\label{th-ansatz}
\varphi = \varepsilon \; \varphi^{(1,0)}(r,\theta) + \varepsilon \; \xi \; \varphi^{(1,1)}(r,\theta) + \varepsilon^2 \; \varphi^{(2,0)}(r,\theta)\,.
\end{align}
By applying the algorithm described earlier, we first solve Eq.~\eqref{eq:constraint}, which describes the evolution of the Chern-Simons scalar $\varphi$.
By putting $V=\varphi$ and by keeping $\mathcal{O}(1,0)$ terms, the evolution equation becomes,
\begin{align}
\label{1st-eq1}
&A \varphi^{(1,0)}_{,rr} + \frac{2} r \varphi^{(1,0)}_{,r} \left( 1 - \frac M r \right) + \frac{1}{r^2} \varphi^{(1,0)}_{,\theta\theta}
+ \frac{\cot{\theta}}{r^2} \varphi^{(1,0)}_{,\theta}-2\frac{d V}{d\varphi}
= - \frac{72 M^3 \mu a_\mathrm{Kerr} }{\nu M r^7} \cos{\theta}\nonumber\\
&\Rightarrow A \varphi^{(1,0)}_{,rr} + \frac{2} r \varphi^{(1,0)}_{,r} \left( 1 - \frac M r \right) + \frac{1}{r^2} \varphi^{(1,0)}_{,\theta\theta}
+ \frac{\cot{\theta}}{r^2} \varphi^{(1,0)}_{,\theta}
= - \frac{72 M^3 \mu a_\mathrm{Kerr} }{\nu M r^7} \cos{\theta}+2\,.
\end{align}
The solution of Eq.~\eqref{1st-eq1} is given by a linear combination of both $\varphi^{(1,0)}_\mathrm{Hom.}$
and $\varphi^{(1,0)}_\mathrm{Part.}$, i.e., $\varphi^{(1,0)} = \varphi^{(1,0)}_\mathrm{Hom.} + \varphi^{(1,0)}_\mathrm{Part.}$.
Because the homogeneous equation is separable,
\begin{align}
\varphi^{(1,0)}_\mathrm{Hom.}(r,\theta) = \varphi(r)_r \varphi_\theta (\theta)+\frac{r(4M+r)}{3}\, .
\end{align}
Thus the homogenous solution of the partial differential equation \eqref{1st-eq1} reduces to ordinary differential equations for $\varphi_r(r)$ and $\varphi_\theta (\theta)$  and have the following solutions,
\begin{align}
\label{Hom-sol-1}
\varphi(r) =&\, c_{1} H\left[\left[\frac{\gamma}{2},\frac{\gamma}{2}\right],\gamma,\frac{2 M} r \right] r^{-\gamma/2}
+ c_{2} H\left[\left[\frac{\gamma_1}{2},\frac{\gamma_1}{2}\right],\gamma_1,\frac{2 M} r \right] r^{-\gamma/2}\, ,
\nonumber \\
\varphi(\theta) =&\, c_3 L\left( -{ \frac{\gamma}{2}}, \cos\theta \right) + c_4 L_1 \left(- {\frac{\gamma}{2}}, \cos{\theta} \right)\, ,
\end{align}
where $H(\cdots)$'s are generalized hypergeometric functions\footnote{
The generalized hypergeometric function $H\left( \left[n_1, n_2,\cdots, n_p \right], \left[d_1, d_2, \cdots, d_q \right], z\right)$ is
generally defined as,
\[
H\left( \bm{n}, \bm{d}, z \right) = \sum_{k=0}^\infty \frac{\prod_{i=1}^p \mathrm{PS} \left( n_i, k \right)}
{\prod_{j=1}^q \mathrm{PS} \left( d_j,k \right)} \frac{z^k}{k!}\, ,
\]
where $\bm{n} = \left[ n_1,n_2,\cdots, n_p \right]$, $\bm{d}= \left[ d_1,d_2,\cdots, d_q \right]$ and $\mathrm{PS}(n,k)$
is the Pochhammer symbol, $\mathrm{PS}(n,k) \equiv \prod_{j=0}^{k-1} \left( n + j \right)$.
$H(\cdots)$'s in (\ref{Hom-sol-1}) correspond to $p=2$ and $q=1$.
},
$L(\cdot)$ is the Legendre polynomial of the first kind\footnote{
The Legendre polynomial of the first kind is defined as,
\[
L(a,z)= H \left( [-a,a+1],[1],\frac{1}{2}(1-z) \right) \, .
\] },
${L_1}(\cdot)$ is the Legendre polynomial of the second kind\footnote{
The Legendre polynomial of the second kind is defined as,
\[
L_1(a,z)=\frac{\sqrt{\pi} \Gamma(1+a) H \left( \left[ 1 + \frac{a}{2}, \frac{1}{2} + \frac{a}{2} \right], \left[ \frac{3}{2}+a \right],\frac{1}{z^2} \right)}
{2 z^{1+a} \Gamma \left( \frac{3}{2}+a \right)2^a}\, .
\] },
$c_1$, $c_2$, $c_3$, and $c_4$ are constants of integration, and the constants $\gamma$ and $\gamma_1$ are given by,
\begin{align}
\label{tilde-alpha}
\gamma = 1 - \sqrt{1 - 4 c_5}\,, \quad
\gamma_1 = 1 + \sqrt{1 - 4 c_5}\, ,
\end{align}
where $c_5$ is a constant of integration accompanied by the separation of variables.

To extract the physical properties of the constants of integration, we study the solution of $\varphi^{(1,0)}$ in detail.
For this aim, we consider the asymptotic behavior of the solution when $r \gg M$, that yields,
\begin{align}
\varphi_r(r) \sim c_{1} \left[ 1 + \frac M{2 r} \gamma \right] r^{-\gamma/2}
+ c_{2} \left[ 1 + \frac M{2 r} \gamma_1\right] r^{-\gamma_1/2}\, .
\end{align}
Because $\varphi$ is a real scalar field, $\gamma$ and $\gamma_1$ should be real numbers, which yields $c_5 < 1/4$.
Additionally, if we require that $\varphi$ gives finite total energy \cite{Nashed:2011fg}, $\varphi$ must decrease into a constant faster than $1/r$,
 that tells $\gamma > 2$ and $\gamma_1>2$.
Because the first condition $\gamma > 2$ cannot be satisfied when $c_5 < 1/4$, we find $c_{1} = 0$,
and the second condition is satisfied if $c_5 < 0$.
By combining all the above discussions, we find
\begin{align}
\varphi^{(1,0)}_\mathrm{Hom.} = \mathrm{const.}
\end{align}

Since we know the form of the homogenous solution of Eq.~\eqref{1st-eq1}, we can now find the particular solution, { up to order $\varepsilon$}, and obtain,
\begin{align}
\label{theta-sol-SR111}
\varphi^{(1,0)}_{_\mathrm{Part.}}(r,\theta)=&\, \frac{{ \varepsilon}}{144 {\nu M^4 r^4}} \left\{72 M^4 r^6\nu \left(1-\frac{37M}{18r}
+ \frac{62 M^2}{9 r^2}\right) \right. \nonumber \\
&\, -192 \left( r-M \right) r^4 \left( M^5\nu+{\frac{45}{64}} a_\mathrm{Kerr} \mu \cos \theta\right) \ln \left( r-2\,M \right) \nonumber\\
&\, \left. +324 \mu \left[ {\frac{5}{12}} r^4 \left( r-M \right) \ln r +\frac{5}{6}Mr^4 \left(1+\frac{M^2}{3r^2}+ \frac{2 M^3}{3 r^3}
+ \frac{6M^4}{5r^4} \right) \right] a_\mathrm{Kerr} \cos\theta \right\}\,,
\end{align}
where we set the extra integration constants equal to zero since they do not have any effect on the modified Einstein equations\footnote{It is important
to stress that the behavior of a scalar field in a background of the Kerr geometry has been investigated when taking into account the axion hair
for the Kerr~\cite{Campbell:1990ai,Reuter:1991cb}, for dyon~\cite{Campbell:1991rz} black holes and cosmology~\cite{Kaloper:1991rw},
and string theory.}. { It is important to stress that the particular solution given by Eq.~\eqref{theta-sol-SR111} cannot reduce to the particular
solution presented in \cite{Yunes:2007ss}.
This is due to  the contribution of the potential $V\propto \varphi$ in the solution \eqref{theta-sol-SR111} which we cannot  isolate from  it.}

{ The behavior of Eq. \eqref{theta-sol-SR111} as $r\to \infty$ is not finite due to the existence of terms like $\ln\,r$, which means that Simon scalar field is not defined as  $r\to \infty$.}  When $r$ is large, Eq.~\eqref{theta-sol-SR111} yields the form:
 \begin{align}
\label{theta-sol-SR11}
{ \varphi^{(1,0)}_{_\mathrm{Part.}}(r,\theta)\approx\, -\frac{4{ \varepsilon} M^2 \cos(\theta) \mu a_\mathrm{Kerr}
}{\nu  r^5}\left(1+\frac{25M}{14r}+\frac{45M^2}{14r^2}+\frac{35M^3}{6r^3}+\frac{32M^4}{3r^4}\right)+\frac{{ \varepsilon} r^2}{2}\left[1+\frac{2M}{3r}(12\ln\,r+37)\right]\,.}
\end{align}
 { The last  term in Eq.~\eqref{theta-sol-SR11} is  mainly comes form the contribution of $V\propto {\varphi}$ and is responsible to   break  the parity symmetry i.e., when  $a_\mathrm{Kerr} \to - a_\mathrm{Kerr}$ one can see that $ {\varphi} \nrightarrow -{\varphi}$.} { It is not easy to find the  horizons of Eq. \eqref{theta-sol-SR11} because the algebraic equation constructed from it is of order 11. }

Since we succeed to derive the Chern-Simons field, we can now try to derive the Chern-Simons corrections of the metric perturbations.
It should be noted that there is a contribution of $\mathcal{O}(2,1)$ from the stress-energy tensor of the Chern-Simons scalar field~\eqref{Tab-theta}
to the equations~\eqref{eom}, and therefore we neglect the contribution to the perturbation of the metric.
In that case, the modified equations~\eqref{eom} are divided into two sets:
The first set is given by a closed system of differential equations including $a^{(1,1)}$, $b^{(1,1)}$, $d^{(1,1)}$, and $f^{(1,1)}$,
which comes from the components $(t,t)$, $(r,r)$, $(r,\theta)$, $(\theta,\theta)$, and $(\phi,\phi)$-components.
The second set of the modified Einstein equations yields a single differential equation for $\Omega^{(1,1)}$, which is the $(t,\phi)$-component.

The first set does not depend on the Chern-Simons field $\varphi$ and thus the contribution of this set vanishes identically.
Therefore, we require to deal with the second set, $(t,\phi)$-component, which yields,
\begin{align}
\label{V1}
0=&\, \frac{\sin\theta }{2 r^5\kappa}\left\{  12\mu Mr^2A {\varphi^{(1,0)}}_{_{r \theta}} +\kappa \sin\theta r^7 A {\Omega^{(1,1)}}_{rr}
+ \sin\theta r^5{\Omega^{(1,1)}}_{\theta \theta} \kappa+6\alpha Mr A {\varphi^{(1,0)}}_{\theta} \right. \nonumber\\
&\, \left. - \left[ 4\kappa \sin \theta r^2 {\Omega^{(1,1)}}_r +3r \kappa {\Omega^{(1,1)}}_\theta \cos \theta
+ \nu \sin\theta \left\{\beta V \left( \varphi \right) - {\varphi^2}^{(1,0)} \right\} \left( \Omega^{(1,1)} r^3 +Ma_\mathrm{Kerr} \right) r^4\right]\right\} \,.
\end{align}
{ Equation~\eqref{V1} reduces to the following form when $V=0$,
\begin{align}
\label{111}
& 2 \sin^2{\theta} \Omega^{(1,1)}_{,\theta\theta} + 3 \sin{2 \theta} \Omega^{(1,1)}_{,\theta} + 8 r A \sin^2{\theta} \Omega^{(1,1)}_{,r}
+2 r^2 A \sin^2{\theta} \Omega^{(1,1)}_{,rr} = \frac{15}{2} \frac{\mu^2}{\nu \kappa} \frac{a_\mathrm{Kerr} A}{r^8} \sin^2{\theta}
\left(3 r^2 + 8 M r + 18 M^2\right)\,,
\end{align}}
{ which coincides with the form derived in Ref. \cite{Yunes:2007ss}.

The general solution of Eq. \eqref{V1} when $V\neq 0$ is again given by a sum of a homogeneous solution and a particular solution.
The particular solution,  up to the leading order in $\varepsilon$ and { $\xi$},  has the following form,}
\begin{align}
\label{w-sol-SR22}
\Omega^{(1,1)} =&\, - \left. \frac{{ \varepsilon\, \xi} \mu^2a_\mathrm{Kerr}}{1792\kappa \nu r^8 M^6} \right\{ 6048 M^8+2240 r^2 M^6-1400 r^5 M^3+11340 r^4 M^4 \nonumber\\
&\, +3360 r^5 M^3 \ln 2 +3840 M^7r-1050 r^6 M^2-1050 r^7M-525 \ln \left(1-\frac{2M} r \right) r^8 \nonumber \\
&\, \left. +6720 \ln \left(1-\frac{2M} r \right) r^5 M^3+5040 \ln \left(1-\frac{2M} r \right) r^4 M^4 \right\}\,.
\end{align}
 { The asymptotic form of Eq.~\eqref{w-sol-SR22} as $r\to \infty$ yields},
\begin{align}
\label{w-sol-SR11}
{\Omega^{(1,1)} \approx \frac{16 { \varepsilon \xi} \mu^2 a_\mathrm{Kerr}}{3 \nu r^6\,\kappa}\frac{M^3}{ r^3}  \left( 1 + \frac{45M}{7r}
+ \frac{405M^2}{154 r^2}+ \frac{35M^3}{8 r^3} + \frac{69M^4}{13 r^4}+ \frac{972M^5}{77r^5} \right)\,.}
\end{align}
%
{ which is  identical with the one derived in Ref. \cite{Yunes:2007ss} when the last three terms are neglected.  The last three terms in Eq. \eqref{w-sol-SR11} are due to the contribution of the potential $V \propto \varphi$. The form of $\Omega^{(1,1)}$ approaches zero as $r\to \infty$.}
The homogeneous solution of Eq.~\eqref{w-sol-SR22} is given by a combination of two independent generalized hypergeometric functions, whose argument is $\frac{r}{2M}$
and the functions include a separation constant $c_6$.
Certain values of such a constant make the solution purely real but at spatial infinity, the solution diverges.
The other values of constant $c_6$ make the solution infinite or complex.
By using the aforementioned discussion, we find that the integration constants, which are coefficients of these hypergeometric functions, must vanish.
Therefore, Eq.~\eqref{w-sol-SR22} represents the full solution.

The full expression of the gravitomagnetic metric perturbation of $\mathcal{O}(\varphi)$ and $\mathcal{O}(\xi)$ has the following form,

\begin{align}
\Omega =&\, \frac{2 { \varepsilon} M a_\mathrm{Kerr}}{r^3} - \left. \frac{{ \varepsilon \xi} \mu^2a_\mathrm{Kerr}}{1792\kappa \nu r^8 M^6}\right\{ 6048 M^8+2240 r^2 M^6-1400 r^5 M^3+11340 r^4 M^4
+3360\, r^5 M^3\ln 2 +3840 M^7r \nonumber\\
&\, +3360 r^5 M^3\ln 2 +3840 M^7 r -1050 r^6 M^2-1050 r^7M-525 \ln \left(1-\frac{2M} r \right) r^8 \nonumber \\
&\, \left. +6720 \ln \left(1-\frac{2M} r \right) r^5 M^3+5040 \ln \left(1-\frac{2 M} r \right) r^4 M^4 \right\}\,,
\label{Omega1}
\end{align}
which asymptotically yields at large $r$ the following form,
\begin{align}
\label{22}
\Omega \approx \frac{2 { \varepsilon} M a_\mathrm{Kerr}}{r^3} -\frac{16{ \varepsilon\,\xi} \mu^2 a_\mathrm{Kerr}}{3 \nu r^6}\frac{M^3}{ r^3} \left( 1 + \frac{45M}{7r}
+ \frac{405M^2}{154 r^2}+ \frac{35M^3}{8 r^3} \right)\,.
\end{align}
%
Eq.~(\ref{22}) represents the first slowly-rotating solution in the dynamical Chern-Simons gravity when the potential does not vanish.
We stress the fact that the perturbation is severely suppressed when $r\to \infty$, decaying like $r^{-4}$, which indicates that its significance is observed in the strong field region.
Now let us discuss solution \eqref{22} with/without the potential $V$. When $V=0$ the solution decaying as $r^{-6}$ and when $V\neq 0$ the solution decaying as $r^{-4}$.
This means that the solution without $V$ decreases faster than the one with $V$.

The parity transformation can be identified with $\Omega\to -\Omega$, which corresponds to $a_\mathrm{Kerr} \to - a_\mathrm{Kerr}$
as found in (\ref{Omega1}).
If the parity symmetry is conserved, the scalar field $\varphi$ is transformed as $\varphi\to -\varphi$
under the transformation $a_\mathrm{Kerr} \to - a_\mathrm{Kerr}$ but as we find in (\ref{theta-sol-SR111}),
$\varphi$ is a non-trivial function of $a_\mathrm{Kerr}$ and under the transformation $a_\mathrm{Kerr} \to - a_\mathrm{Kerr}$,
$\varphi$ is not transformed as $\varphi\to -\varphi$.
This is because the potential $V\propto \varphi$ is the odd function of $\varphi$ and therefore the parity symmetry is explicitly broken.
This situation can be compared with the $V\propto \varphi^2$ in the next subsection, where the model has parity symmetry.

{ By calculating the next order correction to $\varphi$, we can verify that the approximated solution \eqref{theta-sol-SR11} is self-consistent.
The corrections $\varphi^{(2,0)}$ and $\varphi^{(1,1)}$ can be obtained by deriving the solution of the evolution equation to the next order. Doing
so, we find $\varphi^{(2,0)}=0$, and  $\varphi^{(1,1)}$ up to the linear order in $\varepsilon$ and $\xi$  takes the form},
\begin{align}
\label{phi111}
\varphi^{(1,1)}=&\, \frac{{ \varepsilon \xi}}{4233600 \cos\theta r^9 M^8\nu^2} \left\{ 3763200 M^9 r^9\nu^2 \left( M-r \right)
\mathrm{dilog} \left( {\frac r{2M}} \right) -3763200 \left[ - \frac{243}{40} \mu \left( \frac{125}{384} \mu^2 r^5 \left( M-r \right) \ln 2 \right. \right. \right. \nonumber \\
&\, + \left( \frac{25}{144} \nu M^4+ \frac{187}{672} \mu^2 \right) r^6 -\left( \frac{187}{672} \mu^2 M + \frac{25}{144} \nu M^5 \right) r^5
 - \frac{5}{18} \mu^2 r^3 M^3- \frac{5}{9} \mu^2 r^2 M^4 \nonumber \\
&\, \left. \left. -\mu^2r M^5+\mu^2 M^6 \right) a_\mathrm{Kerr} \cos \theta
+ \left( \ln \left( \frac{2M} r \right) \left( M-r \right) +\frac{1}{3} \left( M-2r\right) \right) \nu^2 M^9 r^5 \right] r^4\ln \left( r-2 M \right) \nonumber \\
&\, -3763200\, \left(2 \nu^2 M^9-{\frac{81}{8}}\,\mu^3a_\mathrm{Kerr} \,\cos \theta \right) \left( M-r \right) r^9\ln \left( r-M \right)
 -16934400 \left[ \frac{27}{20} \left( \frac{125}{384} \mu^2 r^5 \left( M-r \right) \ln 2 \right. \right. \nonumber\\
&\, \left. + \left( \frac{25}{144} \nu M^4 + \frac{187}{672} \mu^2 \right) r^6- \left( \frac{187}{672} \mu^2M + \frac{25}{144} \nu M^5 \right) r^5
 - \frac{5}{18} \mu^2 r^3 M^3 - \frac{5}{9} \mu^2 r^2 M^4 - \mu^2r M^5+\mu^2 M^6 \right) r^4 \ln r\nonumber\\
&\, + \frac{135}{128} \left( M^5 + \frac{8}{15} r^5
+ \frac{3}{10} M r^4 + \frac{5}{18} M^3 r^2 + \frac{5}{9} M^4r \right) \mu^2 r^5\ln 2+ \left\{ M^3\left( \frac{279}{1120} \mu^2- \frac{5}{16} \nu M^4 \right) r^6 - \left( \frac{1683}{2240} \mu^2
 \right. \right.\nonumber\\
&\, \left.+ \frac{15}{32} \nu M^4 \right) r^9- M^2\left( \frac{561}{2240} \mu^2+ \frac{5}{32} M^{4 }\nu \right) r^7 + M^4\left( \frac{7551}{5600} \mu^2 - \frac{9}{16} \nu M^4 \right) r^5
+ \frac{1259}{350} M^5\mu^2 r^4 + \frac{125}{196} M^6 \mu^2 r^3 + M^9\mu^2  \nonumber \\
&\,\left.+ \frac{225}{196} M^7\mu^2 r^2 + \frac{5}{4} M^8\mu^2r  \left\} M \left] \mu a_\mathrm{Kerr} \cos \theta
 -5017600 \nu^2 \left[ M \left( M-r \right) \ln \left( r \right) - \frac{27}{128} rM - \frac{57}{128} M^2 + \frac{27}{128} r^2 \right] M^8 r^9 \right\}
\right. \right. \,,
\end{align}
where $\mathrm{dilog}(x)$ is the Dilogarithm function which is defined as $\mathrm{dilog}(x)= \int_1^x dt\, \left(\frac{\ln(t)}{1-t}\right)$.
Eq.~\eqref{phi111} yields the following asymptotic form for small $r$,
\begin{align}
\label{theta_111}
\varphi^{(1,1)}=&\, -{ \frac{25 \mu { \varepsilon \xi}  \cos\theta}{448 \nu M^5} \frac{a_\mathrm{Kerr}}{ r^2}}\left(1 + \frac{2 M} r + \frac{18 M^2}{5 r^2} + \frac{32 M^3}{5 r^3}
+\frac{80 M^4}{7 r^4} + \frac{144 M^5}{7 r^5} + \frac{112 M^6}{5 r^6} + \frac{448 M^7}{25 r^7} \right)\nonumber\\
&\, - \frac{3\,\mu { \varepsilon \,\xi} a_\mathrm{Kerr} \ln \left( {\frac{2M} r} \right)}{2 M^5 r^2{\nu}} \left(1 + \frac{2 M} r + \frac{18 M^2}{5 r^2} + \frac{32 M^3}{5 r^3}\right)
+{ \frac{9M a_\mathrm{Kerr} \mu}{4 r^4\nu}+ \frac{5 a_\mathrm{Kerr} \mu}{4 r^3\nu } + \frac{5 a_\mathrm{Kerr} \mu}{8M r^2\nu } }\,.
\end{align}
The last line in Eq.~\eqref{theta_111} represents the contribution of the potential.
Equation~\eqref{theta_111} is smaller than $\varphi^{(1,0)}$ by a factor $\xi$.
Therefore the small-coupling approximation is surely self-consistent.
By using Eq.~\eqref{theta_111} in the Chern-modified field equation, we obtain a rectification to the metric proportional of $\mathcal{O}\left(\xi^2 \epsilon\right)$,
which we neglect in this study.

\subsection{The case of $ V\propto \varphi^2$}

In this case, the scalar field equation will be affected because $dV=2\varphi$.
By applying Eq.~\eqref{eq:constraint} to the line element~\eqref{slow-rot-ds2} and by using Eqs.~\eqref{cons} and \eqref{th-ansatz}, we obtain
$\mathcal{O}(1,0)$ corrections of the evolution equation in the form,
\begin{align}
\label{1st-eq}
&A \varphi^{(1,0)}_{,rr} + \frac{2} r \varphi^{(1,0)}_{,r} \left( 1 - \frac M r \right) + \frac{1}{r^2} \varphi^{(1,0)}_{,\theta\theta}
+ \frac{\cot{\theta}}{r^2} \varphi^{(1,0)}_{,\theta}-2\frac{d V}{d\varphi}
= - \frac{72 M^2 \mu a_\mathrm{Kerr}}{\nu r^7} \cos{\theta}\nonumber\\
&\Rightarrow\ A \varphi^{(1,0)}_{,rr} + \frac{2} r \varphi^{(1,0)}_{,r} \left( 1 - \frac M r \right) + \frac{1}{r^2} \varphi^{(1,0)}_{,\theta\theta}
+ \frac{\cot{\theta}}{r^2} \varphi^{(1,0)}_{,\theta}-2m^2 \varphi
= - \frac{72 M^2 \mu a_\mathrm{Kerr}}{\nu r^7} \cos{\theta}\,.
\end{align}
Here we wrote $V(\varphi) = m^2 \varphi^2$.
In the following, we may put $m=1$ by redefining $r\to \frac{r}{m}$ and $\nu \to m^5 \nu$.
The homogeneous solution of the above partial differential equation takes the form,
\begin{align}
{ \varphi^{(1,0)}_\mathrm{Hom.}(r,\theta) = \varphi(r)\varphi (\theta)}\,,
\end{align}
where\footnote{{$H_c$ is the Heun Confluent function, $H_c(\alpha,\beta,\gamma,\delta,\eta_1,z)$, which is the solution of the differential equation,
\begin{align}
0= Y''(z)-\frac{ \left[1+\beta-z \left( \alpha-\gamma-\beta-2 \right) -\alpha z^2 \right]Y'(z)}{z(z-1)}
 -\frac{ \left[\alpha \left( \beta+1 \right) - \left[ \left( \beta+\gamma+2 \right) \alpha-2\delta \right]z
 - \left( \gamma+1 \right)\beta-\gamma-2\eta_1 \right]Y(z)}{2z(z-1)}\,, \nonumber
\end{align}
with boundary condition $Y(0)=1$ and $Y'(0)=\frac{\beta \left( 1+\gamma-\alpha \right) +\gamma-\alpha+2\eta_1}{2 \left( \beta+1 \right)}$.
}},
{
\begin{align}
\label{Hom-sol-1B}
\varphi(r) =&\, \e^{2r}H_c\left(8M,0,0,-16M^2,c_7,\frac r{2M}\right)\left[c_{8}+ c_{9}\int\frac{dr}{(2M-r)r\,\e^{4r}\,H_c\left(8M,0,0,-16M^2,c_7,\frac r{2M}\right)}\right]\,,
\nonumber \\
\varphi (\theta) =&\, c_{10} L_1\left(\frac{\alpha}{2},\cos\theta\right) + c_{11} L_2\left(\frac{\alpha}{2},\cos\theta\right)\,.
\end{align}
}
Here $H_c(\cdots)$ is the Heun Confluent function,
$L_i(\cdot)$'s are Legendre functions and associated Legendre functions of the first and second kinds,
$c_7$, $c_8$, $c_9$, $c_{10}$, and $c_{11}$ are constants of integration, and the coefficient $\alpha$ is defined as,
\begin{align}
\label{tilde-alphaB}
\alpha =\sqrt{1 - 4 c_7}\,,
\end{align}
where a constant $c_7$ appears by the separation of variables.
{ For large $r$, the function $\varphi(r)$ takes the form,
\begin{align}
\varphi(r)= c_{12} H\left[\left[\frac{\alpha_1}{2},\frac{\alpha_1}{2}\right],\alpha_1,\frac{2 M} r \right] r^{-\frac{\alpha_1}{2}}
+ c_{13} H\left[\left[\frac{\beta_1}{2},\frac{\beta_1}{2}\right],\beta_1,\frac{2 M} r \right] r^{- \frac{\beta_1}{2}}\,,
\end{align}}
where $H(\cdot)$ is a generalized hypergeometric function, $\alpha_1$ and $\beta_1$ are defined as,
\begin{align}
\label{tilde-alphaC}
\alpha_1 = 1 - \sqrt{1 - 4 c_{14}}\,, \quad \beta_1 = 1 +\sqrt{1 - 4 c_{14}}\,,
\end{align}
where a constant $c_{14}$ also appears by the separation of variables.
%
%
From the above discussion we find that the homogenous solution of Eq. \eqref{1st-eq} is \begin{align} \varphi^{(1,0)}_\mathrm{Hom.} = \mathrm{const.}\end{align}
The particular solution of \eqref{1st-eq}, { up to order $\varepsilon$}, yields the form,
\begin{align}
\label{theta-sol-SR}
\varphi^{(1,0)}{}_{_\mathrm{Part.}}(r,\theta) =&\, \frac{72{ \varepsilon}\cos(\theta) M^2\mu a_\mathrm{Kerr} 
H_c {\e^{\sqrt{2}r}}}{\beta} \nonumber \\
&\, \times \left[ \left( \int \frac{\e^{\sqrt{2}r}H_c} {r^5} dr \right) \left( \int \frac{\e^{-2\,\sqrt{2}r}}{r \left(2 M-r \right) {H_c}^2} dr \right)
 -\int \frac{\e^{\sqrt{2}r}H_c}{r^5} \left( \int \frac{\e^{-2 \sqrt{2}r}}{r \left(2 M-r \right){H_c}^2}{dr}\right) dr \right]\,,
\end{align}
{ where we have put the additional constant of integration to vanish since it has no role in the modification of the Einstein equations.
For small $r$  Eq.~\eqref{theta-sol-SR} yields\footnote{{ Equation \eqref{theta-sol-SR} can be calculated for small $r$ by carried out the integration term by term. }}},
\begin{align}
\label{theta-sol-SR1}
{ \varphi^{(1,0)}{}_\mathrm{Part.}(r,\theta)\approx \frac{5{ \varepsilon}\cos(\theta) \mu}{8\beta M }\frac{a_\mathrm{Kerr}}{r^2}\left(1+\frac{2M} r+\frac{18M^2}{5r^2}\right) \,,}
\end{align}
{ We should note that ${\varphi^{(1,0)}}$ is linear to $a_\mathrm{Kerr}$ and therefore under the transformation $a_\mathrm{Kerr}\to - a_\mathrm{Kerr}$,
which corresponds to the parity transformation, ${\varphi^{(1,0)}}$ is transformed as ${\varphi^{(1,0)}}\to -{\varphi^{(1,0)}}$.
This tells that the parity symmetry is preserved. This is because the potential $V$ is the even function of ${\varphi^{(1,0)}}$. }

Following the procedure of the case $V\propto \varphi$, by calculating ${ \Omega^{(1,1)}}$,
we obtain\footnote{$H_{cp}$ is the HeunCPrime which is the derivative of the Heun Confluent function.},
\begin{align}
\label{V}
& 2 \sin^2{\theta} \Omega^{(1,1)}_{,\theta\theta} + 3 \sin{2 \theta} \Omega^{(1,1)}_{,\theta} + 8 r A \sin^2{\theta} \Omega^{(1,1)}_{,r}
+ 2 r^2 A \sin^2{\theta} \Omega^{(1,1)}_{,rr} +\frac{6\nu\sin^2\theta V(\varphi) \left( Ma_\mathrm{Kerr} +r^3\Omega^{(1,1)} \right)}{r^5}\nonumber\\
&=\frac{216 M^2\mu^2 {\e^{\sqrt{2}r}} a_\mathrm{Kerr} \sin^2 \theta}{r^5{\kappa}{\nu} H_c} \left\{ \e^{2\, \sqrt{2}r} \left(2 M \left( \sqrt{2}r-1 \right) H_c
+H_{cp} r \right) \left( 2M-r \right) H_c \int \frac{ \e^{\sqrt{2}r}H_c}{ r^5}\left[\int \frac{\e^{-2 \sqrt{2}r}}{\left( 2 M-r \right) r {H_c}^2}dr\right]{dr}
\right. \nonumber\\
& \left. \quad - \left( \e^{2\sqrt{2}r} \left(2 M \left( \sqrt{2}r-1 \right) H_c +H_{cp} r \right) \left( 2M-r \right) H_c
\left[\int \frac{\e^{-2 \sqrt{2}r}}{\left( 2M -r \right) r {H_c}^2}dr\right]+2M \right) \int \frac{ \e^{\sqrt{2}r}H_c}{r^5}{dr} \right\}\,.\end{align}
The asymptotic behavior of the right-hand side of Eq.~\eqref{V} yields the following form when $V=0$,
\begin{align}
2 \sin^2 \theta \Omega^{(1,1)}_{,\theta\theta} + 3 \sin{2 \theta} \Omega^{(1,1)}_{,\theta} + 8 r { A} \sin^2{\theta} \Omega^{(1,1)}_{,r}
+2 r^2{ A} \sin^2{\theta} \Omega^{(1,1)}_{,rr} = \frac{15 { A} \mu^2 a_\mathrm{Kerr} }{2 \nu \kappa r^8} \sin^2{\theta} \left(3 r^2 + 8 M r + 18 M^2\right)\,,
\end{align}
which is the identical form derived in \cite{Yunes:2007ss}.
The particular solution of Eq.~\eqref{V}, { up to the linear order of $\varepsilon$ and $\xi$}, is given by
\begin{align}
\label{w-sol-SR1}
\Omega^{(1,1)} =&\, \frac{432{ \varepsilon\, \xi} a_\mathrm{Kerr} \mu^2 M^2}{\left( 2M-r \right){\nu}{\kappa}H_c r^2 } \nonumber \\
&\, \times \int dr\left\{\frac{1}{r^4} \int \left[\left(r-2 M \right) H_c{\e^{\sqrt{2}r}}
\left( 2M \left( \sqrt{2}r-1 \right) H_c+H_{cp} r \right) \int \frac{\e^{\sqrt{2}r}H_c}{r^5} \left[\int \!\frac{\e^{-2 \sqrt{2}r}}{ r\left(2 M-r \right) {H_c}^2}{dr}\right]{dr}
\right. \right. \nonumber\\
&\, \left. \left. +\left\{ \left( 2M-r \right){H_c}{\e^{ \sqrt{2}r}} \left(2 M \left(\sqrt{2}r-1 \right) H_c +H_{cp} r \right)
\int \frac{\e^{-2 \sqrt{2}r}}{ r \left(2 M-r\right) {H_c}^2}{dr}+\frac{2M}{\e^{\sqrt{2}r}} \right\} \int \!\frac{\e^{\sqrt{2}r}H_c }{ r^5}{dr}\right] \right\}\,.
\end{align}
The asymptotic form of Eq.~\eqref{w-sol-SR1} when ${\varphi^{(1,0)}}$ has the form given by Eq.~\eqref{theta-sol-SR1} yields,
\begin{align}
\label{w-sol-SR}
\Omega^{(1,1)} \approx - \frac{5{ \varepsilon\,\xi} { \mu}^2 a_\mathrm{Kerr}}{8 { \nu} \kappa r^6} \left( 1 + \frac{12M}{7r} + \frac{27 M^2}{10 r^2} \right)\,.
\end{align}
which is identical with one derived in \cite{Yunes:2007ss} when the potential $V=0$.

{ To the linear order in { $\xi$} and ${ \varepsilon}$ the full gravitomagnetic metric perturbation is given by
\begin{align}
\Omega^{(1,1)} =&\, \frac{2 { \varepsilon} M a_\mathrm{Kerr}}{r^3} +\frac{432 { \varepsilon \, \xi} a_\mathrm{Kerr} \mu^2 M^2}{\left( 2 M-r \right){\nu}{\kappa}H_c r^2 } \int dr\left\{\frac{1}{r^4} \int \left[\left(r-2 M \right)
H_c{\e^{\sqrt{2}r}} \left( 2M \left( \sqrt{2}r-1 \right) H_c+H_{cp} r \right) \right. \right. \nonumber \\
&\, \times \int \frac{\e^{\sqrt{2}r}H_c}{r^5} \left[\int \!\frac{\e^{-2 \sqrt{2}r}}{ r\left(2\,M-r \right) {H_c}^2}{dr}\right]{dr} \nonumber\\
& \left. \left. +\left( \left( 2M-r \right){H_c}{\e^{ \sqrt{2}r}} \left(2 M \left(\sqrt{2}r-1 \right) H_c +H_{cp} r \right) \int \frac{\e^{-2\,\sqrt{2}r}}{ r \left(2\,M-r\right) {H_c}^2}{dr}
+\frac{2M}{\e^{\sqrt{2}r}} \right) \int \frac{\e^{\sqrt{2}r}H_c }{r^5}{dr} \right] \right\}\,,
\end{align}
which reduces asymptotically to,
\begin{align}
\Omega^{(1,1)} = \frac{2{ \varepsilon}  M a_\mathrm{Kerr}}{r^3} - \frac{5 { \varepsilon \xi}a_\mathrm{Kerr} \xi}{8 r^6} \left( 1 + \frac{12M}{7r} + \frac{27 M^2}{10 r^2} \right)\, .
\end{align}
{ Following the same procedure carried out for  the case of  $V\propto \varphi$ in the last subsection \ref{Vvarphi}, we evaluate $\varphi^{(1,1)}$ by calculating the field equation of the Simon scalar field, Eq. \eqref{eq:constraint}, and get the form $\varphi^{(1,1)}$, up to the linear form of $\varepsilon$ and $\xi$, as,}
\begin{align}
\label{phi11}
\varphi^{(1,1)}=&\, \frac{144 { \varepsilon \,\xi} H_c M^2a_\mathrm{Kerr} \alpha {\e^{2r}}\cos\theta}{ \beta^2{\kappa}}\left\{ \int \left[ \e^{2r}
\left( 72{\alpha}^2M\int \frac{\e^{\sqrt{2}r}}{r^2} \left[ - \left[ 2M \left(\sqrt{2}r-1 \right) H_c
+H_{cp}r \right] \int \frac{\e^{\sqrt{2} r}H_c}{ r^5} \right. \right. \right. \right. \nonumber\\
&\, \times \left[\int \frac{\e^{-2\,\sqrt{2}r}{dr}}{r \left( 2M-r \right) {H_c}^2}\right]{dr} + \left( \left( 2M \left(\sqrt{2}r-1 \right) H_c +H_{cp}r \right)
\int \frac{\e^{-2\sqrt{2}r}}{ r \left( 2M-r \right){H_c}^2}{dr}
+ \frac{2M \e^{-2\sqrt{2}r}}{H_c \left( 2 M-r \right)} \right) \nonumber \\
&\, \left. \times \int \frac{\e^{\sqrt{2}r}H_c}{ r^5}{dr} \right]{dr}
 - r^7 \e^{ \sqrt{2}r} H_c \beta\kappa \left[\int \frac{\e^{\sqrt{ 2}r}H_c}{r^5} \left[\int \frac{\e^{-2\sqrt{2}r}}{r\left(r-2 M\right) {H_c}^2}{dr}\right]{dr} \right. \nonumber \\
& \left. \left. \left. -\int \frac{\e^{\sqrt{ 2}r}H_c}{r^5} {dr}\int \frac{\e^{-2\sqrt{2}r}}{ r \left( 2M-r \right) {H_c}^2}{dr} \right] \right)
\frac{H_c}{r^5}\left(\int \frac{\e^ {-4r}}{r \left(2M-r \right) {H_c}^2}{dr}\right) \right]{dr} \nonumber\\
&\, +\int \frac{\e^{2r} H_c}{r^5} \left[ 72{ \alpha}^2\int \left( \left(2M-r \right) \left(2 M \left( \sqrt{2}r -1\right)H_c +H_{cp} r \right) {\e^{2 \sqrt{2}r}}H_c
\int \frac{\e^{\sqrt{2} r}H_c}{r^5}\left[\int \frac{\e^{-2 \sqrt{2}r}}{r \left( 2 M-r \right){H_c}^2}{dr}\right]{dr} \right. \right. \nonumber\\
&\, \left. - \left( \left(2 M-r \right) \left( 2M \left( \sqrt{2}r-1 \right) H_c +H_{cp} r \right) \e^{2 \sqrt{2}r}H_c
\int \frac{\e^{-2\,\sqrt{2}r}}{r \left(2M-r \right) {H_c}^2}{dr}
+2 M \right) \int \frac{\e^{\sqrt{2}r}H_c}{r^5}{dr} \right) \nonumber \\
&\, \times \frac{M \e^{-\sqrt{2}r{dr}}}{ H_c r^2 \left( 2M-r\right)}
+ r^7{\e^{ \sqrt{2}r}}H_c \beta\kappa \left( \int \frac{\e^{\sqrt{ 2}r}H_c}{r^5} \left[\int \frac{\e^{-2\sqrt{2}r}}{r \left(2M-r \right){H_c}^2}{dr}\right]{dr} \right. \nonumber \\
& \left. \left. \left.
 -\int \frac{\e^{\sqrt{2}r}H_c}{ r^5}{dr}\left[\int \frac{\e^{-2\, \sqrt{2}r}}{r \left(2\,M-r \right){H_c}^2}\right]{dr} \right) \right]{dr}
\int \frac{\e^{-4r}}{r \left(2M-r \right) {H_c}^2}{dr}\right\}\,.
\end{align}
{ Calculating the asymptotic form of Eq. \eqref{phi11} by calculating the integration of each term at small $r$  we get}:
\begin{align}
\label{theta_11}
\varphi^{(1,1)}{ \approx }- {\frac{25\alpha { \varepsilon\, \xi}  \cos\theta }{448 \beta M^5 }\frac{ a_\mathrm{Kerr}}{r^2}} \left(1 + \frac{2 M} r + \frac{18 M^2}{5 r^2} + \frac{32 M^3}{5 r^3}
+\frac{80 M^4}{7 r^4} + \frac{144 M^5}{7 r^5} + \frac{112 M^6}{5 r^6} + \frac{448 M^7}{25 r^7} \right)\,.
\end{align}

\section{Some physical properties of the derived Solutions}\label{properties}

In the following subsections, we discuss some physical properties of the solutions obtained in the previous sections.

\subsection{Line-element}

{ For the case $V\propto \varphi$, the non-vanishing metric components up to the linear order in { $\xi$}  and $\varepsilon$ are},
\begin{align}
\label{sol:metric_elements}
g_{tt} =&\, -A - \frac{2{ \varepsilon^2} {a_\mathrm{Kerr}}^2 M}{r^3} \cos^2{\theta} \, , \nonumber \\
g_{t\phi}=&\, - \frac{2 { \varepsilon} M a_\mathrm{Kerr}} r \sin^2{\theta} - \left\{ \left. \left. \frac{{ \varepsilon \xi} }{1792\kappa \nu r^8 M^6}
\right[\mu^2a_\mathrm{Kerr} \right( 6048 M^8 +2240 r^2 M^6 -1400 r^5 M^3 + 11340 r^4 M^4 \right. \nonumber\\
&\, + 3360 r^5 M^3\ln 2 +3840 M^7r -1050 r^6 M^2-1050 r^7M-525 \ln \left(1-\frac{2\,M} r \right) r^8 \nonumber \\
&\, \left. \left. \left. +6720 \ln \left(1-\frac{2M} r \right) r^5 M^3+5040 \ln \left(1-\frac{2M} r \right) r^4 M^4 \right) \right]\right\}
\sin^2 \theta\, , \nonumber \\
g_{rr} =&\, \frac{1}{A} + \frac{{ \varepsilon^2} {a_\mathrm{Kerr}}^2}{A r^2} \left(\cos^2 \theta - \frac{1}{A} \right)\, , \nonumber \\
g_{\theta \theta} =&\, r^2 + {{ \varepsilon^2} a_\mathrm{Kerr}}^2 \cos^2 \theta\, , \nonumber \\
g_{\phi \phi} =&\, r^2 \sin^2 \theta + {{ \varepsilon^2} a_\mathrm{Kerr}}^2 \sin^2 \theta \left(1 + \frac{2 M} r \sin^2 \theta \right)\, .
\end{align}
{ For the case $V\propto \varphi^2$, the non-vanishing metric components up to the linear order in { $\xi$}  and $\varepsilon$  are},
\begin{align}
\label{sol:metric_elements1}
g_{t\phi} =&\, - \frac{2{ \varepsilon}  M a_\mathrm{Kerr}} r \sin^2{\theta}
 - \left\{ \frac{432 { \varepsilon \,\xi} a_\mathrm{Kerr} \mu^2 M^2}{\left( 2M-r \right){\nu}{\kappa}H_c r^2 }
\int dr \left[ \left. \frac{1}{r^4} \int \right[\left(r-2 M \right) H_c \e^{\sqrt{2}r} \right. \right. \nonumber \\
&\, \times
\left( 2M \left( \sqrt{2}r-1 \right) H_c+H_{cp} r \right) \int \!\frac{\e^{\sqrt{2}r}H_c}{r^5} \left[\int \!\frac{\e^{-2\sqrt{2}r}}{ r\left(2M-r \right) {H_c}^2}{dr}\right]{dr} \\
& \left. \left. \left. +\left( \left( 2M-r \right){H_c}{\e^{\sqrt{2}r}} \left(2 M \left(\sqrt{2}r-1 \right) H_c +H_{cp} r \right) \int \frac{\e^{-2\,\sqrt{2}r}}{ r \left(2 M-r\right)
{H_c}^2}{dr}+\frac{2M}{\e^{\sqrt{2}r}} \right) \int \frac{\e^{\sqrt{2}r}H_c }{r^5}{dr} \right] \right] \right\}\sin^2{\theta}\,, \nonumber
\end{align}
where the other components of Eq.~\eqref{sol:metric_elements1} have the same form as those given by \eqref{sol:metric_elements}.
Both of Eqs.~\eqref{sol:metric_elements} and \eqref{sol:metric_elements1} are exact to $\mathcal{O}(2,0)$, $\mathcal{O}(1,1)$, and $\mathcal{O}(0,2)$.

It is important to stress that the cross terms of the above two Eqs.~\eqref{sol:metric_elements} and \eqref{sol:metric_elements1} cannot be gauged out, i.e.,
there is no coordinate transformation that can remove the cross terms in Eqs.~\eqref{sol:metric_elements} and \eqref{sol:metric_elements1}.

{ Now, let us calculate the Pontryagin density in the case $V\propto \varphi$, up to the linear order in { $\xi$} and $\varepsilon$,  and we obtain},
\begin{align}
\label{inv1}
R_{abcd} \tilde R^{abcd}=\frac{36 \, { \varepsilon}  \left( 15 \mu^2\ln2 -8 M^4\kappa \nu \right) a_\mathrm{Kerr} \cos\theta}
{\nu \kappa M^2 r^7} +\mathcal{O}\left(\frac{1}{r^{13}}\right)\,,
\end{align}
{ and the Pontryagin density in the case $V\propto \varphi^2$, up to the linear order in { $\xi$}  and $\varepsilon$ , yields},
\begin{align}
\label{inv2}
R_{abcd} \tilde R^{abcd} = \frac{288{ \varepsilon}  M^2\,a_\mathrm{Kerr} \cos\theta}{r^7}\,.
\end{align}
The above two equations, Eqs.~\eqref{inv1} and \eqref{inv2} show the Chern-Simons term correction to the order considered in the present study.
Due to Eq.~(\ref{eq:constraint}), the Pontryagin density $R \tilde R $ is shifted by $\square \varphi$ and $\frac{dV}{d\varphi}$,
and its deviation from that in the Kerr spacetime can be calculated from (\ref{phi111}) or (\ref{phi11}).

The new solutions given by Eqs.~\eqref{sol:metric_elements} and \eqref{sol:metric_elements1} modify the inertial frames dragging
generated by the BHs rotations.
This can be done through the angular velocity $\omega_{Z}$ for the observers with vanishing angular momentum, which is defined as,
\begin{align}
\omega_{Z} = -\frac{g_{t\phi}}{g_{\phi\phi}}\,,
\end{align}
which gives, for $V\propto \varphi$,
\begin{align}
\label{ang-mom1}
\omega_{Z} =&\, \frac{2 M { \varepsilon} a_\mathrm{Kerr}}{r^3} - \left. \left. \frac{{ \varepsilon \, \xi}}{1792\kappa \nu r^8 M^6} \right\{ \mu^2a_\mathrm{Kerr}
\right( 6048 M^8 + 2240 r^2 M^6-1400 r^5 M^3 \nonumber \\
&\, +11340 r^4 M^4 + 3360 r^5 M^3 \ln 2 +3840\, M^7r -1050 r^6 M^2-1050 r^7M \nonumber\\
&\left. \left. -525 \ln \left(1-\frac{2M} r \right) r^8 +6720 \ln \left(1-\frac{2M} r \right) r^5 M^3
+5040 \ln \left(1-\frac{2 M} r \right) r^4 M^4 \right) \right\}\,,
\end{align}
and for $V\propto \varphi^2$, yields,
\begin{align}
\label{ang-mom2}
\omega_{Z} =&\, \frac{2 { \varepsilon} M a_\mathrm{Kerr}}{r^3}
+ \frac{432{ \varepsilon\, \xi} a_\mathrm{Kerr} \mu^2 M^2}{\left( 2M-r \right){\nu}{\kappa}H_c r^2 } \int dr \left\{ \left. \frac{1}{r^4} \int \right[ \left(r-2 M \right)
H_c{\e^{\sqrt{2}r}} \right. \nonumber \\
&\, \times \left( 2M \left( \sqrt{2}r-1 \right) H_c+H_{cp} r \right) \int \frac{\e^{\sqrt{2}r}H_c}{r^5} \left[\int \!\frac{\e^{-2 \sqrt{2}r}}{ r\left(2\,M-r \right)
{H_c}^2}{dr}\right]{dr} \nonumber\\
& \left. \left. +\left( \left( 2M-r \right){H_c}{\e^{ \sqrt{2}r}} \left(2 M \left(\sqrt{2}r-1 \right) H_c +H_{cp} r \right) \int \frac{\e^{-2 \sqrt{2}r}}{ r \left(2M-r\right)
{H_c}^2}{dr}+\frac{2M}{\e^{\sqrt{2}r}} \right) \int \frac{\e^{\sqrt{2}r}H_c }{ r^5}{dr} \right] \right\}\,.
\end{align}
Another test to check the physics of the derived solutions for the case $V\propto \varphi$ and $V\propto \varphi^2$
is to study the stability of these solutions using the geodesic deviations.

\section{Geodesic equation}\label{S666g}

To investigate the effect of the slowly rotating BHs solution derived in the previous section,
we study the motion of a test particle of the solution given by Eq.~\eqref{sol:metric_elements1}.
For this aim, we define the worldline $l(\tau)$ of a test particle in a curved space-time
by the Euler-Lagrange equations, which are characterized by,
\begin{align}
\label{4}
\frac{d}{d\tau}\left(\frac{\partial \mathcal{L}}{\partial \dot l^\mu}\right)-\frac{\partial \mathcal{L}}{\partial l^\mu}=0\,,
\end{align}
for the Lagrangian
\begin{align}
\label{7}
2\mathcal{L}=g_{\mu\nu}\dot l^{\mu}\dot l^\nu=-A {\dot{t}^2}
+ \frac{1}{A} {\dot{r}^2}
+ r^2{\dot{\theta}^2}
+ r^2 \sin^2{\theta} \left[ {\dot{\phi}} -\epsilon \Omega(r,\theta) {\dot{t}} \right]^2\,,
\end{align}
where $l^{\mu}(\tau)= \left(t \left(\tau \right), r \left( \tau \right), \theta \left(\tau \right), \phi \left(\tau \right)\right)$
and $\dot l^\mu$ refers to the derivative of $l^\mu$ w.r.t. the affine parameter $\tau$ and $\Omega(r,\theta)$  is given by Eq. \eqref{22}.

We are going to solve the Euler-Lagrange equations (\ref{4}) focusing on the motion at the equatorial plane where $\theta=\pi/2$.
Using this assumption, we obtain the conserved quantities, i.e., the angular momentum $L$ and the energy $E$ as follows,
\begin{align}
\label{E}
E =&\, \frac{\partial \mathcal{L}}{\partial \dot t} = \left(1-\frac{2M}{r}\right)\dot{t}-\epsilon \left(\frac{Ma_\mathrm{Kerr}}{r}+\xi r^2\Omega_1(r)\right)\dot \phi\,,\\
\label{L}
L =&\, \frac{\partial \mathcal{L}}{\partial \dot \phi} = r^2 \dot \phi-\epsilon \left(\frac{Ma_\mathrm{Kerr}}{r}+\xi r^2\Omega_1(r)\right)\dot{t}\,,
\end{align}
where $\Omega_1(r)= -\frac{16\mu^2 a_\mathrm{Kerr}}{3 \nu r^6}\frac{M^3}{ r^3} \left( 1 + \frac{45M}{7r}
+ \frac{405M^2}{154 r^2}+ \frac{35M^3}{8 r^3} \right)$.
Using Eqs.\eqref{E} and \eqref{L} we obtain,
\begin{align}
\label{E1}
\dot{t} =&\,\frac{r(Er^3+\varepsilon L[M a_\mathrm{Kerr}+\xi r^3 \Omega_1(r)])}{r^4-2Mr^3-\varepsilon^2[M^2 a^2_{Kerr}
+2M a_\mathrm{Kerr}\xi r^3\Omega_1(r)+\xi^2r^6\Omega_1{}^2(r)]}\approx \frac{E}{1-\frac{2M}{r}}{ -}\frac{\varepsilon L[\xi\Omega_1(r)
+\frac{M}{r}\frac{a_\mathrm{Kerr}}{r^2}]}{1-\frac{2M}{r}}\,,\nonumber\\
\dot{\phi} =&\,\frac{r(\varepsilon E[M a_\mathrm{Kerr}+\xi r^3 \Omega_1(r)]-2LM)}{r^4-2Mr^3-\varepsilon^2[M^2 a^2_{Kerr}
+2M a_\mathrm{Kerr}\xi r^3\Omega_1(r)+\xi^2r^6\Omega_1{}^2(r)]}\approx \frac{L}{r^2}+\frac{\varepsilon E[\xi\Omega_1(r)+\frac{M}{r}\frac{a_\mathrm{Kerr}}{r^2}]}{1-\frac{2M}{r}}\,.
\end{align}
The use of Eqs.~(\ref{E}) and (\ref{L}), yields an effective potential similar to classical mechanics.
Because $2\mathcal{L}=0$ for the massless particle and $2\mathcal{L}=1$ for the massive particle and by removing
$\dot t$ and $\dot \phi$ through the use of Eqs.~(\ref{E}) and (\ref{L}), and by supposing $\theta=\pi/2$, we obtain to the leading order
\begin{align}
\label{177}
\frac{E^2}{\left(1-\frac{2M}{r}\right)}-\frac{1}{\left(1-\frac{2M}{r}\right)}\dot r^2 - \frac{L^2}{r^2}
+\frac{\varepsilon EL[\xi\Omega_1(r)+\frac{M}{r}\frac{a_\mathrm{Kerr}}{r^2}]}{\left(1-\frac{2M}{r}\right)} = \sigma\,,
\end{align}
with $\sigma = 0$ for massless particles and $\sigma = 1$ for massive particles.
We rewrite Eq.~(\ref{177}) as follows,
\begin{align}
\label{cons1}
0 = \frac{1}{2}\dot r^2 - \frac{1}{2} E^2 -\frac{1}{2}\varepsilon EL\left[\xi\Omega_1(r)+\frac{M}{r}\frac{a_\mathrm{Kerr}}{r^2}\right]+ \frac{1}{2} \frac{L^2}{r^2}\left(1-\frac{2M}{r}\right)
+ \frac{1}{2} \sigma \left(1-\frac{2M}{r}\right) \,.
\end{align}
From Eq.~\eqref{cons1}, we obtain the effective potential $\mathcal{V}(r)$ as,
\begin{align}
\label{eq:pot11}
\mathcal{V}(r) =&\, \frac{1 }{2}\left(1-\frac{2M}{r}\right) \left(\frac{L^2}{r^2}+ \sigma\right) - \frac{1}{2} E^2
 -\frac{1}{2}\varepsilon EL\left[\xi\Omega_1(r)+\frac{M}{r}\frac{a_\mathrm{Kerr}}{r^2}\right] \, .
\end{align}
Using Eq.~\eqref{eq:pot11} in \eqref{cons1}, we find,
\begin{align}
\label{EffE}
\frac{1}{2}\dot r^2 + \mathcal{V}(r)=0\,.
\end{align}
For the study of the perihelion shift, we reparametrize $r(\tau)$ as $r(\phi)$, which yields,
\begin{align}
\label{eq:rphi1}
\frac{1}{2}\frac{\dot r^2}{\dot \phi^2} + \frac{1}{\dot \phi^2}\mathcal{V}(r) = \frac{1}{2}\left(\frac{dr}{d\phi}\right)^2 + \frac{1}{\dot \phi^2}\mathcal{V}(r)=0\,,
\end{align}
where $\dot \phi$ is given by Eq.~\eqref{E1}.
For circular timelike orbits $\sigma = 1$ for a massive particle, it is also possible to solve the equations $\mathcal{V} = 0$ and $\mathcal{V}' = 0$.
The obtained expressions are, however, not so insightful.
{ We consider a perturbation around a circular orbit $r= r_\mathrm{crc.}$ and by plugging in the ansatz $r(\phi)= r_\mathrm{crc.} + r_\phi(\phi)$ in Eq.~(\ref{eq:rphi1}), we obtain $\left(\frac{d r_\phi}{d\phi}\right)^2$, up to the linear order in { $\xi$}  and $\varepsilon$, as,}
\begin{align}
\left(\frac{d r_\phi}{d\phi}\right)^2 \approx - 2\left[ \frac{(r_\mathrm{crc.} + r_\phi)^4}{{L^2}}+\Psi(r_\mathrm{crc.} + r_\phi)\right] \mathcal{V}\left( r_\mathrm{crc.} + r_\phi \right)\,,
\end{align}
{ where $\Psi(r(\phi))=\Psi(r_\mathrm{crc.} + r_\phi)$  can be constructed from Eq.~\eqref{E1} and takes the form:
\begin{align}
\Psi(r(\phi))=-2\frac{(r_\mathrm{crc.} + r_\phi)^6}{L^3}\frac{ E[\xi\Omega_1((r_\mathrm{crc.} + r_\phi))+\frac{M}{(r_\mathrm{crc.} + r_\phi)}\frac{a_\mathrm{Kerr}}{(r_\mathrm{crc.} + r_\phi)^2}]}{1-\frac{2M}{(r_\mathrm{crc.} + r_\phi)}}\,,
\end{align}
where $\Omega_1((r_\mathrm{crc.} + r_\phi))$ is defined after Eq. \eqref{E}. The explicate form of   $L(r_\mathrm{crc.} + r_\phi)$ and $E(r_\mathrm{crc.} + r_\phi)$  are given in Appendix.}
Assuming that the ratio $r_\phi/r_c$ is small, the right-hand side can be expanded into powers of this parameter to second order
\begin{align}
\left(\frac{d r_\phi}{d\phi}\right)^2
\approx - \left[\frac{r_\mathrm{crc.}^4}{{ L^2}}+ \varepsilon \Psi\left({r_\mathrm{crc.}}\right)\right] \mathcal{V}'' \left( r_\mathrm{crc.} \right){r_\phi}^2 \, ,
\end{align}
where we use the fact that $V \left(r_\mathrm{crc.} \right) = 0$ and $V' \left(r_\mathrm{crc.} \right)=0$ for circular orbits, as discussed above.
The above equation, which represents a simple harmonic oscillation, shows that the solution of $r_\phi$ oscillates with a wave number
$K = \sqrt{\left[\frac{r_\mathrm{crc.}^4}{{ L^2}}+ \varepsilon \Psi\left({r_\mathrm{crc.}}\right)\right] \mathcal{V}'' \left(r_\mathrm{crc.} \right)}$ and thus the perihelion shift is given as,
\begin{align}
\Delta \phi =2\pi\left(\frac{1}{K}-1\right) 
\end{align}
Now, we obtain the explicit form of the perihelion shift of massive objects with the potential $\mathcal{V}$ and $\sigma=1$.
We calculate the equations $\mathcal{V} \left( r_\mathrm{crc.} \right) = 0$ and $\mathcal{V}' \left(r_\mathrm{crc.} \right) = 0$
with $L= L_0 + \epsilon\, L_1$ and $E = E_0 +\epsilon\, E_1$.
Through the use of the above data, we can obtain the zeroth order of $L_0 \left(r_\mathrm{crc.} \right)$ and $E_0 \left(r_\mathrm{crc.} \right)$
which are lengthy and we find that the contribution comes from $\varepsilon$ is very weaker w.r.t. the solution of Schwarzschild solution
which ensures that the same feature can be expected in the rotation curves of galaxies.

\section{Stability of the BHs given by Eqs.~(\ref{sol:metric_elements}) and (\ref{sol:metric_elements1}) using geodesic deviation}\label{S666}

The trajectories of a test particle in a gravitational field are described by the geodesic equations which have the form,
\begin{equation}
\label{ge}
\frac{d^2 x^\alpha}{ d\epsilon^2} + \left\{ \begin{array}{c} \alpha \\ \mu \nu \end{array} \right\}
\frac{d x^\mu}{d\epsilon} \frac{d x^\nu}{d\epsilon}=0 \, ,
\end{equation}
with $\epsilon$ being the affine parameter along the geodesic.
The geodesic deviation takes the form \cite{dInverno:1992gxs},
\begin{equation}
\label{ged}
\frac{d^2 \zeta^\alpha}{d\epsilon^2} + 2\left\{ \begin{array}{c} \alpha \\ \mu \nu \end{array} \right\}
\frac{d x^\mu}{d\epsilon} \frac{d \zeta^\nu}{d\epsilon} + \left\{ \begin{array}{c} \alpha \\ \mu \nu \end{array} \right\}_{,\rho}
\frac{d x^\mu}{d\epsilon} \frac{d x^\nu}{d\epsilon}\zeta^\rho=0 \, ,
 \end{equation}
with $\zeta^\rho$ being the deviation 4-vector.
Applying (\ref{ge}) and (\ref{ged}) into \eqref{sol:metric_elements},
we obtain the geodesic equations in the following form,
\begin{align}
\frac{d^2 t}{d\epsilon^2}+\varepsilon \left[ Ma_\mathrm{Kerr} +r^3\xi \Omega \right]\left(\frac{d \phi}{d\epsilon}\right)^2\left\{1-\frac{A'}{2r}\right\}=0\, ,
\quad \frac{1}{2}A'(r)\left( \frac{dt}{d\epsilon}\right)^2 -r\left(\frac{d\phi}{d\epsilon}\right)^2=0\, ,\quad
\frac{d^2\theta}{d\epsilon^2}=0\, , \quad \frac{d^2 \phi}{d\epsilon^2}=0\, .
\end{align}
Using the circular orbit
\begin{equation}
\theta = \frac{\pi}{2}\, , \quad \frac{d\theta}{d\epsilon}=0\, , \quad \frac{dr}{d\epsilon}=0\, ,
\end{equation}
the geodesic deviation is obtained in the following form,
\begin{align}
\label{ged22}
0= &\, \frac{d^2 \zeta^0}{d\epsilon^2} +\frac{A'}{A}\frac{dt}{d\epsilon} \frac{d\zeta^1}{d\epsilon}
+\varepsilon \left[ Ma_\mathrm{Kerr} +r^3\xi \Omega \right]\left(2\frac{d \phi}{d\epsilon} \frac{d \zeta^3}{d\epsilon}-\frac{A'} r\frac{dt}{d\epsilon} \frac{d \zeta^0}{d\epsilon}\right)
+\left\{\varepsilon\xi r^2 \left[ 3\Omega+r \Omega' \right] \left( \frac{d\phi}{d\epsilon}\right)^2 \right.\nonumber \\
&\left. -\frac{\varepsilon}{2r^2} \left[ 2\xi r^3A'\Omega+r^4A'\xi\Omega'-Ma_\mathrm{Kerr} A'+A''Ma_\mathrm{Kerr} r+\xi A''r^4\Omega \right]
\left( \frac{dt}{d\epsilon}\right)^2 \right\}\zeta^1 \, , \nonumber\\
0= &\, 2 \frac{d^2 \zeta^1}{d\epsilon^2}-2Ar \frac{d\phi}{d\epsilon}\frac{d\zeta^3}{d\epsilon}+2AA' \frac{dt}{d\epsilon}\frac{d\zeta^0}{d\epsilon}
 -\left\{2 \left[ A+rA' \right]\left( \frac{d\phi}{d\epsilon}\right)^2 - \left[ AA''+A'^2 \right]\left( \frac{dt}{d\epsilon}\right)^2\right\}\zeta^1 \,,\nonumber\\
0= &\, \frac{d^2 \zeta^2}{d\epsilon^2} + \left( \frac{d\phi}{d\epsilon} \right)^2 \zeta^2\,, \quad
0=\frac{d^2 \xi^3}{d\epsilon^2} +\frac{2} r \frac{d\phi}{d\epsilon}
\frac{d\xi^1}{d\epsilon} \, ,
\end{align}
where the functions $A=1-\frac{2M} r$ and $\Omega$ is defined either by the second term of the $g_{t \phi}$ of Eq.~(\ref{sol:metric_elements}) or Eq.~(\ref{sol:metric_elements1}).

The third equation of (\ref{ged22}) represents a simple harmonic motion, which ensures that the motion in
the plan $\theta=\frac{\pi}{2}$ is stable.
Assuming the solutions of the remaining equations of Eq.~(\ref{ged22}) to be in the form,
\begin{align}
\label{ged33}
\zeta^0 = n_1 \e^{i \omega \phi}\, , \quad \zeta^1= n_2\e^{i \omega \phi}\, ,
\quad \mbox{and} \quad \zeta^3 = n_3 \e^{i \omega \phi}\, ,
\end{align}
where
$n_1$, $n_2$, $n_3$, and $\omega$ are constants.
Substituting (\ref{ged33}) into (\ref{ged22}), we obtain $\omega$ to the leading order of $\varepsilon$ as,
\begin{align}
\label{con1}
\omega=&\, \frac{1}{ \sqrt{4A - 2A' r} \left(6 A^2 - A'' A - \left( 3- r^2 \right) A'^2-5A' A r \right) r^2}
\left\{ \sqrt{2} r^2 \left( 6 A'^2- A'' A - \left( 3- r^2 \right)A'^2-5A' A r \right)^{3/2} \right. \nonumber\\
&\, +\epsilon A' \left[ Ma_\mathrm{Kerr} r\left( Ma_\mathrm{Kerr} +\xi r^3\Omega \right) \left[ AA'' -2 A'^2 \right]+A \left[ \xi r^4\left(1- r^2 \right) \Omega'
+ \xi r^3 \left( 2+ r^2 \right) \Omega-Ma_\mathrm{Kerr} \right. \right. \nonumber \\
&\, \left. \left. \left. +4 r^2Ma_\mathrm{Kerr} \right] A'
 -2r \left(\xi r^3 \left[ \Omega - r\Omega' \right] \right) A^2 \right] \right\} +\mathcal{O}\left(\varepsilon^2\right)\,.
\end{align}
The stability condition for the spacetimes (\ref{sol:metric_elements}) or (\ref{sol:metric_elements1}) is the positivity of $\omega$ i.e., $\omega>0$ \cite{Misner:1973prb}.
Equation~(\ref{con1}) coincides with that derived in \cite{Nashed:2020mnp} when $\varepsilon=0$.
We draw the condition (\ref{con1}) in Figure~\ref{Fig:1} for the two cases,
$V\propto \varphi$ and $V\propto \varphi^2$ using Eqs.~(\ref{sol:metric_elements}) or (\ref{sol:metric_elements1}). { As for the case $V\propto {\varphi}$ it is important to stress that the value of the parameter $\mu$ is the key role to make the BH stable or not. For example when $0.01<\mu$ we have always unstable BH otherwise, i.e., $0<\mu\leq0.01$ we have stable BH. Same discussion can be used for the case $V\propto {\varphi}^2$ but in this case to get stable BH we must have $0<\mu\leq0.001$ and if $\mu>0.001$ we get unstable BH. The preceding discussion shows that the weaken of the parameter $\mu$ is necessary to create a stable BH as clear for the case $V\propto {\varphi}^2$  which is not true for the case $V\propto {\varphi}$. Finally, in the previous discussion we fixed all the parameters of the model and variate the parameter $\mu$. The same discussion can be applied if we fix all the parameters of the model and variate the parameter $\nu$. For the variation of the parameter $\nu$ and for the case $V\propto \varphi$, all the numerical values used before for the parameters of the model are the same except that we put $\mu=0.01$.  In this case we should have $\nu\geq 0.01$ to get a stable model and for the case $V\propto \varphi^2$  we should have $\nu\geq 0.001$ to get a stable model provided that all the numerical values used before for the parameters of the model are the same except that we put the value of the parameter $\mu=0.001$. Same discussion can be carried out if we variate the rotation parameter $a_{Kerr}$ and fixed all the other parameters. }
\begin{figure}
\centering
\subfigure[~The plot of $\omega$ with respect to the radial coordinate $r$ when $V\propto \varphi^2$. Here we assume the numerical values of
$M$,  $\nu$ $\cdots$ as $M=0.7$, $\nu=0.01$, $\varepsilon=1$ and $a_{Kerr}=0.1$, { $\theta=\pi/2$}.]{\label{fig:2a}\includegraphics[scale=0.3]{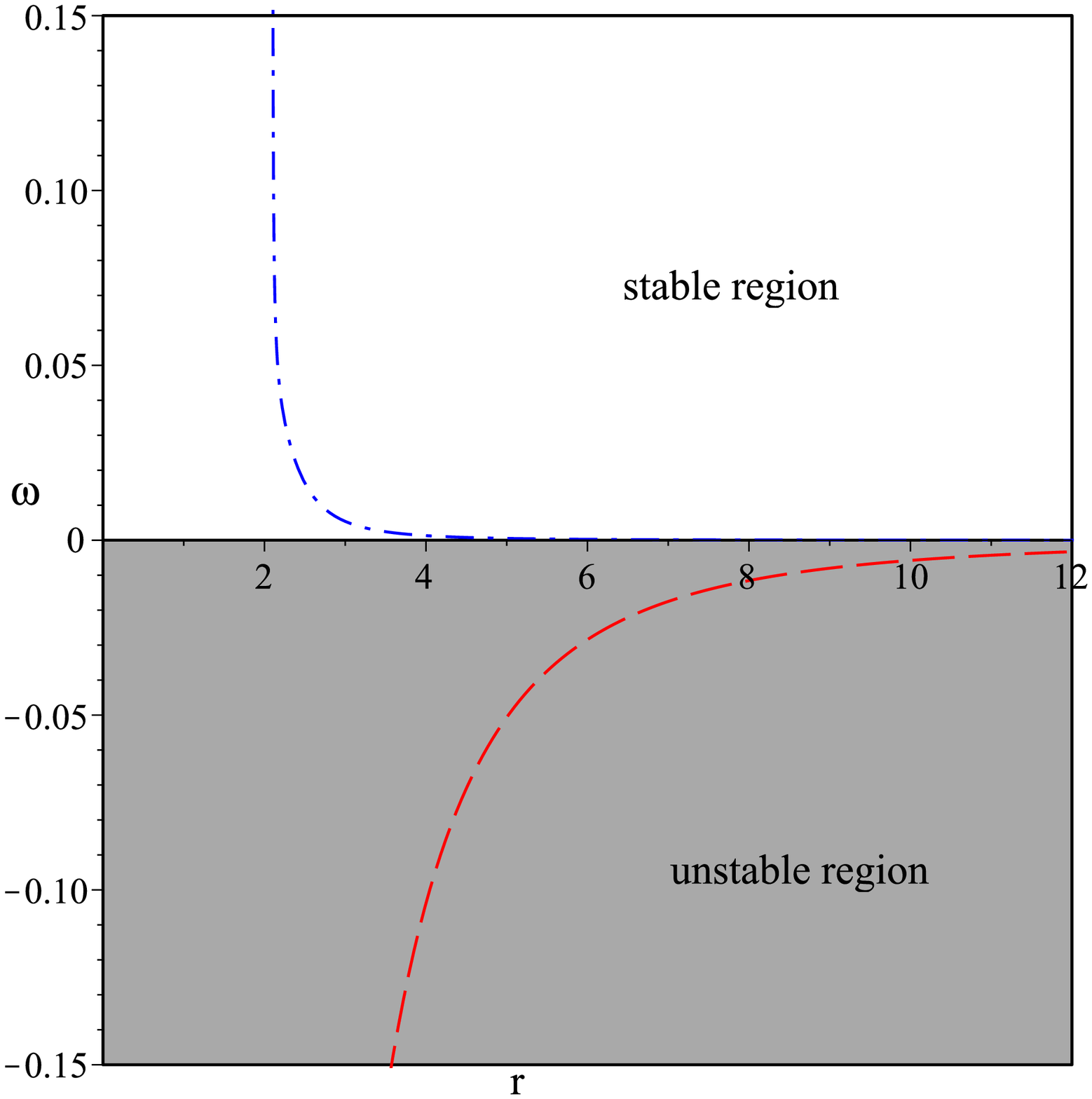}}\hspace{0.6cm}
\subfigure[~The plot of $\omega$ with respect to the radial coordinate $r$ when $V\propto \varphi^2$. Here we assume the numerical values of
$M$,  $\nu$ $\cdots$ as $M=0.01$, $\nu=0.001$, $\varepsilon=0.0001$ and $a_{Kerr}=0.01$, { $\theta=\pi/2$}.]{\label{fig:2b}\includegraphics[scale=0.3]{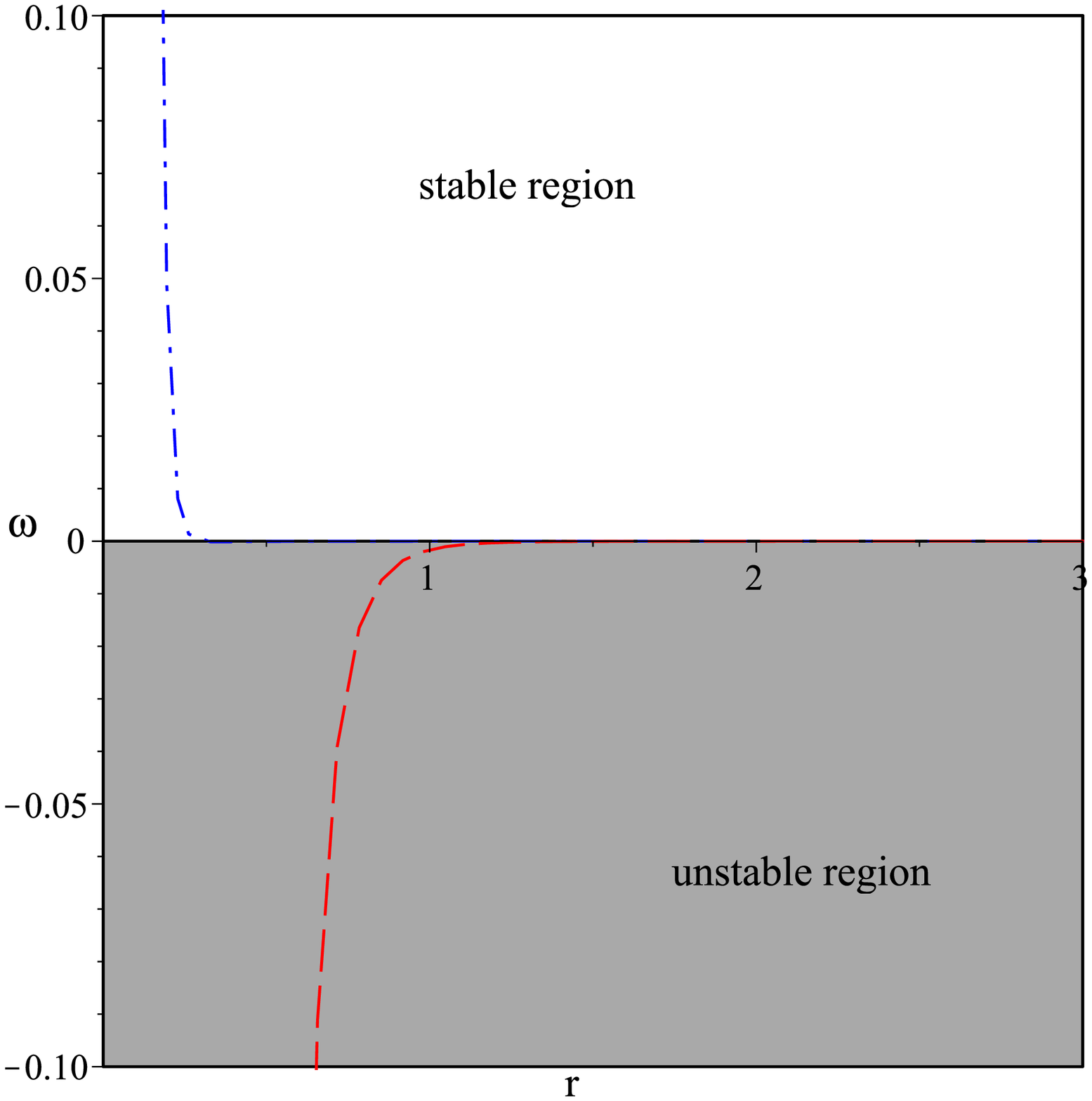}}
\caption{Schematic plots of the $\omega$ with respect to the radial coordinate $r$ of the black hole solutions given by Eqs.~(\ref{sol:metric_elements}) and (\ref{sol:metric_elements1}).}
\label{Fig:1}
\end{figure}

\section{Conclusion and discussions}
\label{conclusions}

Yunes and Pretorius have derived a non-trivial slowly rotating solution in the frame of the dynamical Chern-Simons modified gravity
setting the potential of this theory equal to zero \cite{Yunes:2007ss}.
In this study, we extend their solution by taking into account the potential and assuming it to have three different forms.
We did not tackle the non-dynamical case because the result of this case is not changed from those presented in \cite{Yunes:2007ss}.
Return our attention to the dynamical case where we take the potential of either $V=\mathrm{const.}$ or $V\propto \varphi$ or $V\propto \varphi^2$.

For the case, $V=\mathrm{const.}$, the scalar Chern-Simons field, \eqref{eq:constraint}, did not affect by this choice because $dV=0$
and therefore the form of the scalar field, $\varphi$ coincides completely with what was derived in \cite{Yunes:2007ss}.
However, the field equations \eqref{eom} are affected and despite this affection, the correction of the metric, $\Omega^{(1,1)}$, did not change
and therefore, this case did not supply any new physics different from that presented in \cite{Yunes:2007ss}.

As for the case $V\propto \varphi$, the Chern-Simons correction for the scalar field equation \eqref{eq:constraint} is affected, and as a consequence,
the form of the scalar field $\varphi^{(1,0)}$ and $\varphi^{(1,1)}$ are different from those presented in \cite{Yunes:2007ss}.
Also when we used the form of the scalar field $\varphi^{(1,0)}$ into the field equation \eqref{eom}, we obtained the correction of the metric to $\mathcal{O}\left(\varepsilon\right)$.
This correction, $\Omega^{(1,1)}$, yields an asymptotic form of order $\frac{1}{r^4}$ which is much stronger
than the one presented in \cite{Yunes:2007ss} whose leading order was $\left(\frac{1}{r^6}\right)$.
This case is very interesting because the energy-momentum components of the scalar field have terms of $\mathcal{O}\left(\varepsilon\right)$,
which came mainly from the contribution of the potential.
This is in contrast to the results presented in \cite{Yunes:2007ss} because the energy-momentum components of the scalar field were of $\mathcal{O}\left(\varepsilon^2\right)$.
We have shown that the leading term of the invariant $R_{\nu\mu\alpha\beta} \tilde R^{\mu\nu\alpha\beta}$ is of order $\left(\frac{1}{r^7}\right)$, which is consistent
with the results presented in the literature although the second leading term is of order $\left(\frac{1}{r^{13}}\right)$, which is stronger than that presented in the literature \cite{Yagi:2012ya}.

As for the case $V\propto \varphi^2$, the Chern-Simons correction of the scalar field equation, \eqref{eq:constraint}, is affected, and therefore,
the scalars $\varphi^{(1,0)}$ and $\varphi^{(1,1)}$ are affected.
However, the field equations \eqref{eom} are not effected because $V\propto \varphi^2=\mathcal{O}\left(\varepsilon^2\right)$.
Despite this, the solution of the field equations which yields $\Omega^{(1,1)}$ is different from the one presented in \cite{Yunes:2007ss} because in this case the form
of the scalar field $\varphi^{(1,0)}$ plays the main role in this solution.
The asymptotic behavior of $\Omega^{(1,1)}$ coincides exactly with the one presented in \cite{Yunes:2007ss} without any additional terms.
It is of interest to stress the fact that the components of the energy-momentum tensor of them will be of order $\mathcal{O}\left(\varepsilon^2\right)$
and therefore all the energy conditions are satisfied.
Additionally, we show that the invariant $R_{abcd} \tilde R^{abcd}$ is of order $\left(\frac{1}{r^7}\right)$.

Finally, we discussed the stability of these solutions in the cases $V\propto \varphi$ and $V\propto \varphi^2$ using the geodesic deviations.
We derive the conditions of the stability analytically and discuss them by showing their behaviors graphically.

We conclude our discussion with the following comment.
In this study, we assumed the potential to be either $V\propto \varphi$ or $V\propto \varphi^2$.
Other higher-order case, i.e., $V\propto \varphi^n$ and $n>2$, is not permitted because in that case all the solutions either of the scalar field
or the correction of the metric will be of order $\varepsilon^{n_1}$ $\left(n_1>1\right)$.
Another case that deserves study is to assume $V\propto \varphi^{n_2}$ and in that case $n_2$ is a fraction.
This case will be studied elsewhere.

\section*{Appendix}
The explicate form of $L(r_\mathrm{crc.} + r_\phi)$ and $E(r_\mathrm{crc.} + r_\phi)$   are given as follows:
Since the potential is given by Eq.~\eqref{eq:pot11} and
we will consider the case of massless, i.e., $\left(\sigma=0\right)$.
Then we get
\begin{align}
\label{eq:pot11B}
\mathcal{V}'(r) =&\, - \frac{L^2}{r^3}\left( 1 - \frac{3M}{r}\right)
 -\frac{1}{2}\varepsilon EL\left[\xi\Omega'(r) - \frac{3Ma_\mathrm{Kerr}}{r^4} \right] \, ,
\end{align}
where
\begin{align}
\label{asymOmegaB}
\Omega'_1 (r)= - \frac{6M a_\mathrm{Kerr}}{r^4} + \frac{5 a_\mathrm{Kerr} \xi}{8 r^7} \left( 6 + \frac{84M}{r} + \frac{108 M^2}{5 r^2} \right)\, .
\end{align}
When $\varepsilon=0$, by using the equations $\mathcal{V}(r) =\mathcal{V}'(r) =0$, we obtain
\begin{align}
\label{calc1}
r=r_0 \equiv 3M\, , \quad E= E_0 \equiv - \frac{\left| L \right|}{3\sqrt{3} M}\, ,
\end{align}
which expresses the standard photon sphere.
We also note that the binding energy $E$ should be negative, $E<0$.
By assuming
\begin{align}
\label{calc2}
r= r_0 + \varepsilon r_1 \, , \qquad \qquad E=E_0 + \varepsilon E_1\, ,
\end{align}
and by using Eq.~(\ref{eq:pot11}) with $\sigma=0$ and Eq.~(\ref{eq:pot11B}), the equations $\mathcal{V}(r) =\mathcal{V}'(r) =0$ give,
\begin{align}
\label{calc3}
0 =&\, - E_0 E_1 -\frac{1}{2} E_0 L\left[\xi\Omega \left(r_0 \right)+\frac{Ma_\mathrm{Kerr}}{27M^3}\right] \, , \quad
\Omega \left( r_0 \right) = \frac{2 a_\mathrm{Kerr}}{27 M^2} - \frac{53 a_\mathrm{Kerr} \xi}{11664 M^6} \, , \\
\label{calc4}
0 =&\, - \frac{L^2}{9 M^4} r_1 -\frac{1}{2} E_0 L\left[\xi\Omega' \left(r_0 \right) - \frac{Ma_\mathrm{Kerr}}{27 M^4} \right] \, , \quad
\Omega' \left( r_0 \right) = - \frac{2 a_\mathrm{Kerr}}{27 M^3} + \frac{91 a_\mathrm{Kerr} \xi}{87484 M^7} \, .
\end{align}
Using Eqs.~\eqref{calc3} and ~\eqref{calc4}, we obtain
\begin{align}
\label{calc5}
r=r_\mathrm{circ.} \equiv &\, 3M - \frac{3 \varepsilon \mathrm{sign} ( L ) M^3}{2\sqrt{3}}\left[\xi\Omega' \left(r_0 \right) - \frac{Ma_\mathrm{Kerr}}{27 M^4} \right] \, , \\
\label{calc6}
E=&\, - \frac{\left| L \right|}{3\sqrt{3} M}  -\frac{1}{2} L\left[\xi\Omega \left(r_0 \right)+\frac{Ma_\mathrm{Kerr}}{27M^3}\right] \, .
\end{align}
Here $\mathrm{sign} (x)$ is defined by
\begin{align}
\label{calc7}
\mathrm{sign} (x) \equiv \left\{ \begin{array}{cc}
+1 &\ \mbox{when}\ x\geq 1 \\
 -1 &\ \mbox{when}\ x< 1
\end{array} \right. \, .
\end{align}

\section*{Acknowledgments}
The authors thank the anonymous referee for constructive criticism that improves the manuscript's presentation.

\end{document}